\documentclass[12pt,preprint]{aastex}
\usepackage{natbib}
\usepackage{hyperref}
\usepackage{lineno}

\usepackage{subfigure}

\usepackage{lipsum}

\newcommand{\Fermi}{{\em Fermi}}
\newcommand{\Swift}{{\em Swift}}

\shorttitle{\Fermi~LAT Stacking Analysis of \Swift~Localized Gamma-ray Bursts}
\shortauthors{The \Fermi~LAT and \Fermi~GBM Collaborations}

\begin{document}

\title{\Fermi~LAT Stacking Analysis of \Swift~Localized GRBs}

\author{
M.~Ackermann\altaffilmark{1}, 
M.~Ajello\altaffilmark{2}, 
B.~Anderson\altaffilmark{3}, 
W.~B.~Atwood\altaffilmark{4}, 
M.~Axelsson\altaffilmark{5,6}, 
L.~Baldini\altaffilmark{7}, 
G.~Barbiellini\altaffilmark{8,9}, 
D.~Bastieri\altaffilmark{10,11}, 
R.~Bellazzini\altaffilmark{12}, 
P.~N.~Bhat\altaffilmark{13}
% ...
}

\begin{abstract}

We perform a comprehensive stacking analysis of data collected by the \Fermi~Large Area Telescope (LAT) of gamma-ray bursts (GRB) localized by the \Swift~spacecraft, which were not detected by the LAT but which fell within the instrument's field of view at the time of trigger.  We examine a total of 79 GRBs by comparing the observed counts over a range of time intervals to that expected from designated background orbits, as well as by using a joint likelihood technique to model the expected distribution of stacked counts.  We find strong evidence for subthreshold emission at MeV to GeV energies using both techniques.  This observed excess is detected during intervals that include and exceed the durations typically characterizing the prompt emission observed at keV energies and lasts at least 2700 s after the co-aligned burst trigger.  By utilizing a novel cumulative likelihood analysis, we find that although a burst's prompt gamma-ray and afterglow X-ray flux both correlate with the strength of the subthreshold emission, the X-ray afterglow flux measured by {\it Swift's} X-ray Telescope (XRT) at 11 hr post trigger correlates far more significantly. Overall, the extended nature of the subthreshold emission and its connection to the burst's afterglow brightness lend further support to the external forward shock origin of the late-time emission detected by the LAT.  These results suggest that the extended high-energy emission observed by the LAT may be a relatively common feature but remains undetected in a majority of bursts owing to instrumental threshold effects. 

\end{abstract}

\keywords{Gamma-rays: Bursts: Prompt}

\altaffiltext{1}{Deutsches Elektronen Synchrotron DESY, D-15738 Zeuthen, Germany}
\altaffiltext{2}{Department of Physics and Astronomy, Clemson University, Kinard Lab of Physics, Clemson, SC 29634-0978, USA}
\altaffiltext{3}{Royal Swedish Academy of Sciences Research Fellow, funded by a grant from the K. A. Wallenberg Foundation}
\altaffiltext{4}{Santa Cruz Institute for Particle Physics, Department of Physics and Department of Astronomy and Astrophysics, University of California at Santa Cruz, Santa Cruz, CA 95064, USA}
\altaffiltext{5}{Department of Physics, KTH Royal Institute of Technology, AlbaNova, SE-106 91 Stockholm, Sweden}
\altaffiltext{6}{Tokyo Metropolitan University, Department of Physics, Minami-osawa 1-1, Hachioji, Tokyo 192-0397, Japan}
\altaffiltext{7}{Universit\`a di Pisa and Istituto Nazionale di Fisica Nucleare, Sezione di Pisa I-56127 Pisa, Italy}
\altaffiltext{8}{Istituto Nazionale di Fisica Nucleare, Sezione di Trieste, I-34127 Trieste, Italy}
\altaffiltext{9}{Dipartimento di Fisica, Universit\`a di Trieste, I-34127 Trieste, Italy}
\altaffiltext{10}{Istituto Nazionale di Fisica Nucleare, Sezione di Padova, I-35131 Padova, Italy}
\altaffiltext{11}{Dipartimento di Fisica e Astronomia ``G. Galilei'', Universit\`a di Padova, I-35131 Padova, Italy}
\altaffiltext{12}{Istituto Nazionale di Fisica Nucleare, Sezione di Pisa, I-56127 Pisa, Italy}
\altaffiltext{13}{Center for Space Plasma and Aeronomic Research (CSPAR), University of Alabama in Huntsville, Huntsville, AL 35899, USA}
\altaffiltext{14}{Istituto Nazionale di Fisica Nucleare, Sezione di Bari, 70126 Bari, Italy}
\altaffiltext{15}{Istituto Nazionale di Fisica Nucleare, Sezione di Torino, I-10125 Torino, Italy}
\altaffiltext{16}{Dipartimento di Fisica Generale ``Amadeo Avogadro" , Universit\`a degli Studi di Torino, I-10125 Torino,c Italy}
\altaffiltext{17}{W. W. Hansen Experimental Physics Laboratory, Kavli Institute for Particle Astrophysics and Cosmology, Department of Physics and SLAC National Accelerator Laboratory, Stanford University, Stanford, CA 94305, USA}
\altaffiltext{18}{NASA Goddard Space Flight Center, Greenbelt, MD 20771, USA}
\altaffiltext{19}{Laboratoire Univers et Particules de Montpellier, Universit\'e Montpellier, CNRS/IN2P3, Montpellier, France}
\altaffiltext{20}{Laboratoire Leprince-Ringuet, \'Ecole polytechnique, CNRS/IN2P3, Palaiseau, France}
\altaffiltext{21}{Consorzio Interuniversitario per la Fisica Spaziale (CIFS), I-10133 Torino, Italy}
\altaffiltext{22}{INAF-Istituto di Astrofisica Spaziale e Fisica Cosmica, I-20133 Milano, Italy}
\altaffiltext{23}{Istituto Nazionale di Fisica Nucleare, Sezione di Perugia, I-06123 Perugia, Italy}
\altaffiltext{24}{Dipartimento di Fisica, Universit\`a degli Studi di Perugia, I-06123 Perugia, Italy}
\altaffiltext{25}{College of Science, George Mason University, Fairfax, VA 22030, resident at Naval Research Laboratory, Washington, DC 20375, USA}
\altaffiltext{26}{email: jchiang@slac.stanford.edu}
\altaffiltext{27}{Agenzia Spaziale Italiana (ASI) Science Data Center, I-00133 Roma, Italy}
\altaffiltext{28}{INAF Osservatorio Astronomico di Roma, I-00040 Monte Porzio Catone (Roma), Italy}
\altaffiltext{29}{Department of Physics, Stockholm University, AlbaNova, SE-106 91 Stockholm, Sweden}
\altaffiltext{30}{The Oskar Klein Centre for Cosmoparticle Physics, AlbaNova, SE-106 91 Stockholm, Sweden}
\altaffiltext{31}{The Royal Swedish Academy of Sciences, Box 50005, SE-104 05 Stockholm, Sweden}
\altaffiltext{32}{INAF Istituto di Radioastronomia, 40129 Bologna, Italy}
\altaffiltext{33}{Dipartimento di Astronomia, Universit\`a di Bologna, I-40127 Bologna, Italy}
\altaffiltext{34}{Dipartimento di Fisica, Universit\`a di Udine and Istituto Nazionale di Fisica Nucleare, Sezione di Trieste, Gruppo Collegato di Udine, I-33100 Udine}
\altaffiltext{35}{Universit\`a Telematica Pegaso, Piazza Trieste e Trento, 48, 80132 Napoli, Italy}
\altaffiltext{36}{Universit\`a di Udine, I-33100 Udine, Italy}
\altaffiltext{37}{Dipartimento di Fisica ``M. Merlin" dell'Universit\`a e del Politecnico di Bari, I-70126 Bari, Italy}
\altaffiltext{38}{Erlangen Centre for Astroparticle Physics, D-91058 Erlangen, Germany}
\altaffiltext{39}{Laboratoire AIM, CEA-IRFU/CNRS/Universit\'e Paris Diderot, Service d'Astrophysique, CEA Saclay, 91191 Gif sur Yvette, France}
\altaffiltext{40}{Space Science Division, Naval Research Laboratory, Washington, DC 20375-5352, USA}
\altaffiltext{41}{NASA Postdoctoral Program Fellow, USA}
\altaffiltext{42}{Department of Physics and Center for Space Sciences and Technology, University of Maryland Baltimore County, Baltimore, MD 21250, USA}
\altaffiltext{43}{Center for Research and Exploration in Space Science and Technology (CRESST) and NASA Goddard Space Flight Center, Greenbelt, MD 20771, USA}
\altaffiltext{44}{School of Physics and Astronomy, University of Southampton, Highfield, Southampton, SO17 1BJ, UK}
\altaffiltext{45}{Funded by a Marie Curie IOF, FP7/2007-2013 - Grant agreement no. 275861}
\altaffiltext{46}{Science Institute, University of Iceland, IS-107 Reykjavik, Iceland}
\altaffiltext{47}{email: daniel.kocevski@nasa.gov}
\altaffiltext{48}{The George Washington University, Department of Physics, 725 21st St, NW, Washington, DC 20052, USA}
\altaffiltext{49}{Department of Astronomy, Stockholm University, SE-106 91 Stockholm, Sweden}
\altaffiltext{50}{Institute of Space Sciences (IEEC-CSIC), Campus UAB, E-08193 Barcelona, Spain}
\altaffiltext{51}{Department of Physics and Department of Astronomy, University of Maryland, College Park, MD 20742, USA}
\altaffiltext{52}{Hiroshima Astrophysical Science Center, Hiroshima University, Higashi-Hiroshima, Hiroshima 739-8526, Japan}
\altaffiltext{53}{Istituto Nazionale di Fisica Nucleare, Sezione di Roma ``Tor Vergata", I-00133 Roma, Italy}
\altaffiltext{54}{Center for Cosmology, Physics and Astronomy Department, University of California, Irvine, CA 92697-2575, USA}
\altaffiltext{55}{Instituto de Astronomia, Geof\'isica e Cincias Atmosf\'ericas, Universidade de S\~{a}o Paulo, Rua do Mat\~{a}o, 1226, S\~{a}o Paulo - SP 05508-090, Brazil}
\altaffiltext{56}{Department of Physical Sciences, Hiroshima University, Higashi-Hiroshima, Hiroshima 739-8526, Japan}
\altaffiltext{57}{Max-Planck-Institut f\"ur Physik, D-80805 M\"unchen, Germany}
\altaffiltext{58}{email: judith.racusin@nasa.gov}
\altaffiltext{59}{Funded by contract FIRB-2012-RBFR12PM1F from the Italian Ministry of Education, University and Research (MIUR)}
\altaffiltext{60}{Institut f\"ur Astro- und Teilchenphysik and Institut f\"ur Theoretische Physik, Leopold-Franzens-Universit\"at Innsbruck, A-6020 Innsbruck, Austria}
\altaffiltext{61}{National Research Council Research Associate, National Academy of Sciences, Washington, DC 20001, resident at Naval Research Laboratory, Washington, DC 20375, USA}
\altaffiltext{62}{NYCB Real-Time Computing Inc., Lattingtown, NY 11560-1025, USA}
\altaffiltext{63}{Instituci\'o Catalana de Recerca i Estudis Avan\c{c}ats (ICREA), Barcelona, Spain}
\altaffiltext{64}{Max-Planck Institut f\"ur extraterrestrische Physik, 85748 Garching, Germany}

\section{Introduction}
 
Observations by the {\it Fermi Gamma-ray Space Telescope}, with its unprecedented energy coverage, have led to an exciting and productive period for gamma-ray burst (GRB) astronomy.  The {\it Fermi} Gamma-ray Burst Monitor (GBM) has detected over 1500 GRBs in 5 yr of operations, with over 90 of these bursts detected by the \Fermi~Large Area Telescope (LAT) above 40 MeV\footnote{\url{http://fermi.gsfc.nasa.gov/ssc/observations/types/grbs/lat_grbs/}}. The high-energy emission observed by the LAT is typically longer lasting and delayed in onset in comparison to emission at keV energies detected by the GBM.

The origin of the high-energy emission observed by the LAT has been much debated within the GRB community.  Emission detected contemporaneous by the GBM and LAT during the prompt phase of several bursts (e.g. GRB~090217, GRB~090323, and GRB~130427A) suggests that some of the emission at MeV energies observed by the LAT may originate from a simple extension of the prompt GRB spectra into the LAT energy range \citep{Ackermann2013}.  At the same time, the delayed onset and long-lived nature of the LAT emission has led to speculation  \citep{Kumar2009, Ghisellini2010, DePasquale2010, Razzaque2010} that the afterglow components commonly observed at optical and X-ray wavelengths may also produce a significant amount of gamma-ray emission from the high-energy extension of the synchrotron spectrum of the external forward shock.  Broadband fits to the simultaneous multiwavelength observations of GRB~110731A \citep{GRB110731A} and GRB~130427A \citep{GRB130427A_LAT}, which were detected by both \Swift~and \Fermi, show remarkably similar late-time spectral and temporal behavior, lending support to an external shock origin of their late-time GeV emission.  In the case of GRB 130427A, the GeV emission persisted for almost a day after the burst and matched the temporal decay slope observed in the X-ray and optical wavelengths. Moreover, spectral and temporal {\it NuSTAR} observations taken 1.5 and 5 days after the event onset, combined with {\it Swift, Fermi}, and ground-based optical data, unambiguously establish a single afterglow spectral component from optical to multi-GeV energies, most certainly due to synchrotron radiation \citep{GRB130427A_NuSTAR}.  Therefore, there is now a growing body of evidence suggesting that the high-energy emission observed by the LAT is likely due to a combination of both prompt and afterglow contributions in the LAT energy window. 

Despite this converging picture as to the origin of the LAT-detected emission, there still remains a large number of bright bursts for which no high-energy emission was observed by the LAT.  Of the over 1500 bursts detected by the GBM in the 8 keV to 40 MeV energy range in the first 5 yr of operation, only 8$\%$ of the bursts that have occurred within the LAT field of view (FOV) were detected above 40 MeV \citep{Ackermann2013}.  An examination by \citet{UpperLimitsCatalog} of the prompt emission from bright GBM-detected bursts that fell within the LAT FOV, but were not detected by the instrument, showed that many of these bursts either require spectral breaks or have intrinsically steeper prompt spectra than inferred from fitting the GBM data alone in order to explain their nondetections by the LAT.  Likewise, \citet{GRB130427A_NuSTAR} showed that multiwavelength data collected for GRB~130427A were best fit by a smoothly broken power law 1.5 days (and possibly up to 5 days) after the burst.  These observations indicate that spectral breaks may persist between the {\it Swift's} X-ray Telescope (XRT) and LAT windows for thousands of seconds after the prompt emission. 

In this paper, we present a stacking analysis of LAT data of {\it Swift}-detected GRBs that fell within the LAT FOV at the time of trigger, in order to search for subthreshold emission at MeV and GeV energies.  By using well-localized \Swift~bursts, we can eliminate any ambiguity that arises from the positional uncertainty of the burst in the LAT FOV.  We examine a total of 79 GRBs by comparing the observed counts over a range of time intervals to that expected from designated background orbits, as well as by using a joint likelihood method to model the expected distribution of counts, in order to search for subthreshold emission. We find a significant excess above the combined background during time intervals including and exceeding the durations characterizing the prompt emission observed at keV energies using both methods. This analysis follows a similar study by \citet{Lange13}, who performed a counting analysis of GBM-detected bursts with no detectable emission in the LAT above 100 MeV.  In their study, the authors found similar evidence for extended subthreshold emission.  Here we use the joint likelihood analysis to obtain a robust estimate of the flux and spectral properties of this emission.

The paper is structured as follows: In \S\ref{sec:InstrumentOverview}, we review the characteristics of the \Fermi~GBM, \Fermi~LAT, \Swift~BAT, and \Swift~XRT instruments.  In \S\ref{sec:SampleDefinition}, we define the GRB samples considered in this work, and we outline the analysis performed in \S\ref{sec:Analysis}.  We present the results in \S\ref{sec:Results} and discuss the implications of our results in \S\ref{sec:Discussion}.
 
\section{Instrument Overview } \label{sec:InstrumentOverview}
\subsection{{\it Fermi~LAT and Fermi~GBM} } \label{sec:FermiInstrumentOverview}

The {\it Fermi Gamma-ray Space Telescope} consists of two primary instruments, the GBM and the LAT. The GBM has 14 scintillation detectors that together view the entire unocculted sky. Triggering and localization are performed using 12 sodium iodide (NaI) detectors with different orientations placed around the spacecraft in four clusters of three. Two bismuth germanate (BGO) scintillators are placed on opposite sides of the spacecraft so that at least one detector can see any triggered event. GBM spectroscopy uses both the NaI and BGO detectors, sensitive between 8 keV and 1 MeV and between 150 keV and 40 MeV, respectively.  Together they provide nearly 4 decades of energy for unprecedented sensitive spectroscopic studies of GRBs \citep{Meegan:09}. 

The LAT is a pair conversion telescope comprising a $4\times4$ array of silicon strip trackers and cesium iodide (CsI) calorimeters covered by a segmented anti-coincidence detector to reject charged-particle background events. The LAT covers the energy range from 20\,MeV to more than 300\,GeV with an FOV of $\sim 2.4$ sr. The deadtime of the LAT is nominally 26\,$\mu$s, which is crucial for observations of high-intensity transient events such as GRBs.  The LAT triggers on many more background events than celestial gamma-rays. Onboard background rejection is supplemented on the ground using event class selections that accommodate the broad range of sources of interest \citep{Atwood:09}.

\subsection{{\it Swift~BAT and Swift~XRT}} \label{sec:SwiftInstrumentOverview}

The \Swift~spacecraft consists of the Burst Alert Telescope (BAT; \citealt{Barthelmy05}), (XRT) \citep{Roming05}, and Optical and Ultraviolet Telescope (UVOT; \citealt{Roming05}). The BAT is a wide-field, solid-state gamma-ray detector, covering an FOV of 1.4 sr and an imaging energy range of 15--150 keV. The instrument's coded mask allows for positional accuracy of 1$'$ to 4$'$ within seconds of the burst trigger. The XRT is a focusing X-ray telescope covering an energy range from 0.2--10 keV and providing a typical localization accuracy of $\sim3"$.  The UVOT is a clear-aperture Ritchey-Chretien telescope that provides optical and ultra-violet photometry and subarcsecond positional accuracy of the long-lived afterglow counterparts to the prompt emission from GRBs.

\newcommand{\Tzero}{\mbox{$T_0$}}

\begin{figure}
\includegraphics[width=1\columnwidth]{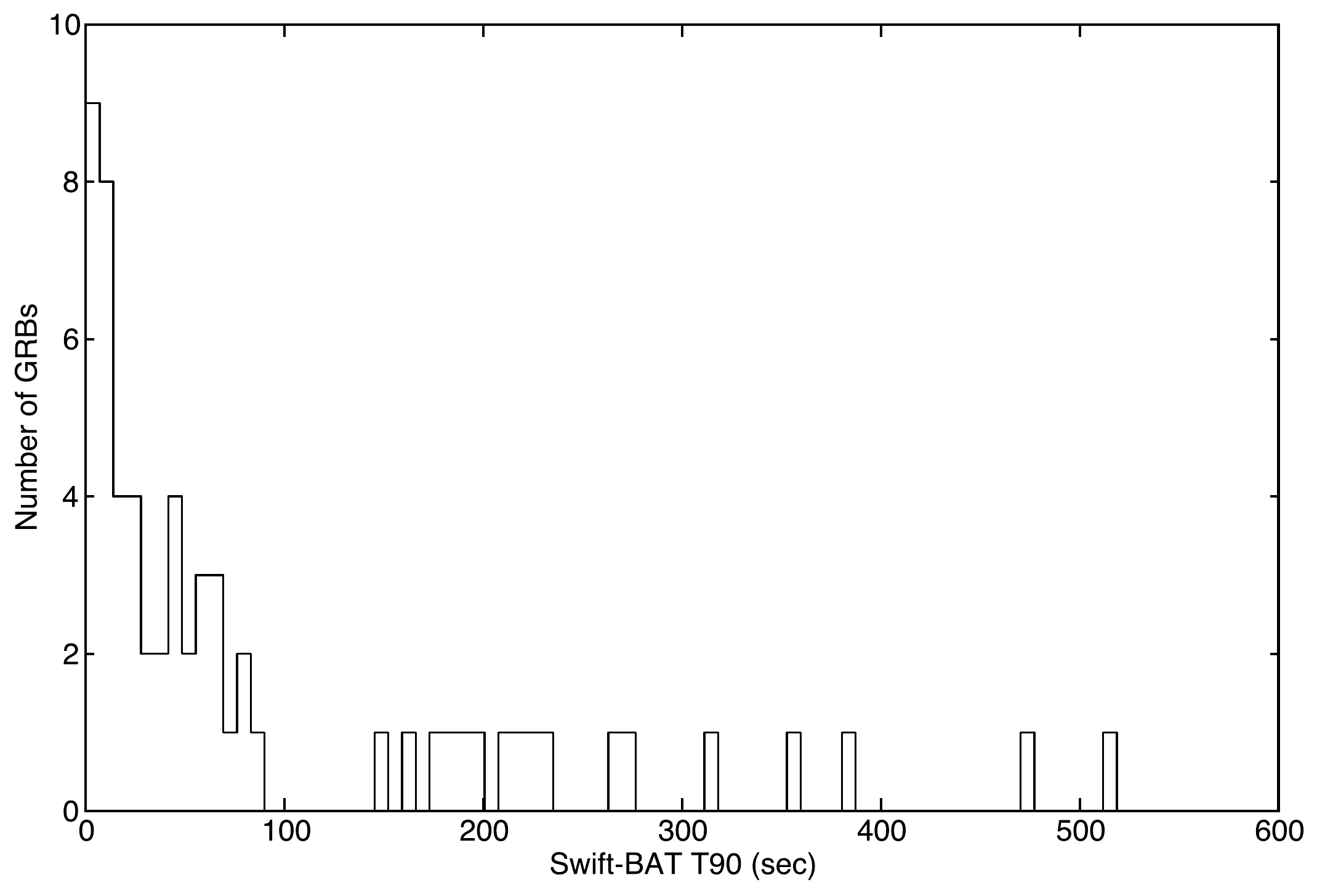}
\caption{The $T_{90}$ duration distribution of the subsample of bursts for which prompt Swift BAT observations were available.  The median duration is 48.9 s, with most bursts having a duration below 100 s}
\label{Fig:T90Distribution}
\end{figure}

\section{Sample Definition} \label{sec:SampleDefinition}

We compiled a sample of all GRBs observed by \Swift~between the beginning of \Fermi~science operations on 2008 August 4 and 2012 February 1, reported by \citet{Donato2012}, yielding a total of 369 GRBs.  Of these, 121 bursts fell within 65$^\circ$ of the LAT z-axis (or boresight), which we define as the LAT FOV, at the time of trigger.  The sensitivity of the LAT falls as a function of off-axis angles; therefore, bursts detected at angles greater than $65^\circ$ were not considered for this analysis.  Of these 121 bursts, we excluded 16 GRBs that were detected by the LAT above 40 MeV and an additional 25 GRBs that occurred during spacecraft passages through the South Atlantic Anomaly (SAA) or at angles with respect to the Earth's zenith that were $\ga 105^\circ$, placing the burst near the Earth's horizon.  Observations of such high zenith angles result in emission at the burst location that is dominated by gamma-rays from the Earth's limb produced by interactions of cosmic rays with the Earth's atmosphere.  The remaining sample includes 79 GRBs. 

Of the 79 GRBs in this sample, 68 were originally detected by BAT.  The remaining 11 bursts were detected by other spacecraft and triggered subsequent follow-up observations by XRT.  Of the total sample, 13 GRBs obtained their best localization through BAT detections, 28 through XRT detections, 18 through UVOT detections, 16 through ground-based followup observations, and 4 from other missions, yielding a median positional uncertainty of 120 arcsec, 1.6$''$, 0.$''$6, and 0.$''$5 respectively \citep{Donato2012}.  All of these positional uncertainties are far smaller than the LAT point spread function (PSF), which has a $68\%$ containment  (i.e., the radius of the circle containing 95$\%$ of the PSF) that varies from $5^\circ$ at 100 MeV to 0.1$^{\circ}$ at 100 GeV with the `P7REP\_SOURCE\_V15' instrument response functions.  Finally, eight of these GRBs were detected by GBM and triggered an autonomous repoint request (ARR) of the \Fermi~spacecraft and another six bursts occurred while the spacecraft was performing a target of opportunity (ToO) observation.  Of the bursts with prompt BAT observations, the median $T_{90}$ duration is 48.9 s.  The $T_{90}$ distribution of this subsample of bursts is shown in Figure \ref{Fig:T90Distribution} for reference.

The complete list of the bursts in our final sample, their positions, and positional uncertainty is given in Table 1.

\section{Analysis} \label{sec:Analysis} 

\subsection{Counting Analysis Overview} \label{sec:CountingAnalysis}

The analysis presented here focuses on two distinct techniques with which to search for a signal in stacked data.  The first method takes the sum of the observed counts collected over a specific duration, energy range, and region of interest (ROI) on the sky centered on each burst's best known position and compares it to the counts collected over an interval deemed to adequately represent the expected background at the time of trigger.  The significance of the stacked counts compared to the stacked background is then estimated through Gaussian statistics.  

The size of the ROI is typically chosen to reflect the 95\% containment radius of the LAT energy-dependent PSF at a particular energy.  For this analysis, we examine both a fixed 10$^\circ$ and 12$^\circ$ energy-independent ROI for each burst, as well as an energy-dependent ROI with a size ranging from 12$^\circ$ at 100 MeV to 2.2$^\circ$ at 10 GeV.  The size of the energy-dependent ROI reflects the 95$\%$ containment of the LAT PSF when considering the P7REP\_SOURCE\_V15 instrument response functions.  From these ROIs, we select `Source' class events that occurred within the first 2700 s after each GRB's trigger time ($T_{0}$) with an energy between 75 MeV and 30 GeV.  The ``Source" data class was specifically optimized for the study of point-like sources, with stricter cuts against non-photon-background contamination in comparison to the ``Transient" data class that is typically used to study GRBs on very short timescales \citep{LATPerformancePaper}.  We also excluded events with an estimated angle from the Earth's zenith greater than 105$^\circ$ to guard against possible contamination from photons that originate from the Earth's limb.  The selection of 75 MeV as the minimum energy is motivated by the possibility of detecting emission from bursts with significant attenuation emission below 100 MeV, (e.g.,~\citealt{Ackermann2013}).   The choice of 2700 s reflects the longest amount of time that most sources on the sky can continuously stay in the LAT FOV before being occulted by the Earth (the exception being those located near the north orbital pole).  

Each of our ROI selections has its own set of advantages and disadvantages.  The energy-dependent ROI has the advantage of collecting photons within a radius that is commensurate with the LAT 95\% containment radius at a given energy, potentially reducing background contamination at high energies, where the LAT PSF is narrowest.  The 12$^\circ$ energy-independent ROI, on the other hand, makes no assumptions regarding the exact shape of the LAT PSF, at the cost of collecting more background at higher energies.  The 10$^\circ$ energy-independent ROI is an intermediate solution, encompassing less of the PSF tail at low energies, but still collecting higher levels of background at high energies compared to the energy-dependent ROI.  The use of a 10$^\circ$ ROI also reflects the better PSF of the \emph{Source} class selection used in this analysis, compared to the \emph{Transient} class selection, for which a 12$^\circ$ ROI is typically used.

\begin{figure*}[t]
\begin{center}
\includegraphics[width=5.35in]{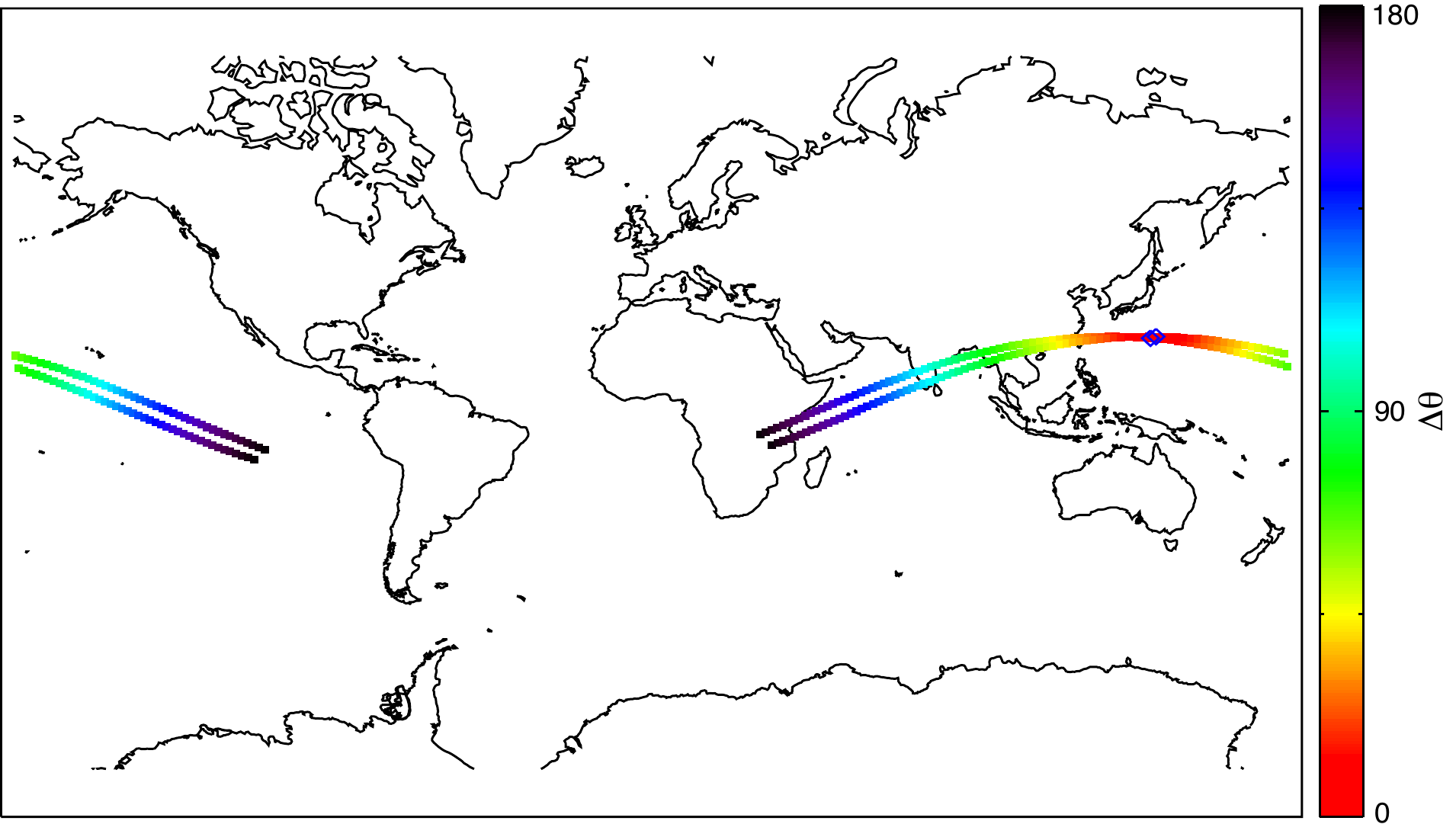}
\end{center}
\caption{A comparison of the orbits from which the signal and background counts are extracted.  The blue diamonds correspond to the location of the spacecraft at the time of the burst trigger and its location exactly 30 sidereal orbits prior.  The color of the orbits represents the angle from the burst position on the sky to the instrument boresight.}
\label{Fig:OrbitalComparison}
\end{figure*}

\subsection{Joint Likelihood Analysis Overview} \label{sec:CompositeLikelihoodAnalysis}

The second technique we employ to search for a subthreshold signal in the stacked LAT data consists of a joint likelihood analysis using the analysis tools developed by the LAT team (ScienceTools version v9r30p1).\footnote{{\tt http://fermi.gsfc.nasa.gov/ssc/}}  In standard unbinned likelihood fitting of individual sources, the expected distribution of counts for each burst is modeled as a point source using an energy-dependent LAT PSF and a power-law source spectrum with a normalization and photon index that are left to vary as free parameters.  Galactic and isotropic background components are also included in this model.  The Galactic component, \emph{gll\_iem\_v05}, is a spatial and spectral template that accounts for interstellar diffuse gamma-ray emission from the Milky Way.  The normalization of the Galactic component is kept fixed during the fit.  The isotropic component, \emph{iso\_source\_v05}, provides a spectral template to account for all remaining isotropic emission that is not represented in the Galactic diffuse component and therefore accounts for contributions from both residual charged-particle backgrounds and the isotropic time-averaged celestial gamma-ray emission. The normalization of the isotropic component is allowed to vary during the fit.  Both the Galactic and isotropic templates are publicly available\footnote{http://fermi.gsfc.nasa.gov/ssc/data/access/lat/BackgroundModels.html}.

We can then derive the probability density of the observed data given this model and create a ``likelihood" function by treating this probability density as a function of the model parameters.  The likelihood function is essentially the probability of observing the data given the chosen model for a range of model parameters.  By maximizing the likelihood function with respect to our parameters of interest, we can then estimate the model parameters that make the observed data the most probable.

The maximum likelihood technique can be expanded to apply to multiple sources to constrain or estimate a set of parameters thought to be common to those sources.  In this case, the backgrounds and properties of each source can be modeled individually in source-specific likelihoods, as described above. The source-specific likelihoods can then be multiplied to yield a joint likelihood function that is used for inference on the common parameter of interest.  In this way, the joint likelihood becomes a source stacking technique, which has been previously applied to LAT data in the search for dark matter (Ackermann et al. 2011) and to study the extragalactic background light (Ackermann et al. 2012b).  This method is implemented in the \Fermi~Science Tools as the \emph{Composite2} routine.

To quantify the significance of a potential excess above background, we employ the common likelihood-ratio test \citep{Neyman1928}.  For this test, we form a test statistic (TS) that is the ratio of the likelihood evaluated at the best-fit parameters under a background-only, null hypothesis, i.e. a model that does not include a point-source component, to the likelihood evaluated at the best-fit model parameters when including a candidate point source at the center of the ROI.  According to Wilks's theorem, this ratio is distributed approximately as $\chi^{2}$, so we choose to reject the null hypothesis when the TS is greater than TS $= 25$, roughly equivalent to a $5 \sigma$ rejection criterion. 

The data selection for this method is identical to that performed for the counting analysis in the case of the 10$^\circ$ and 12$^\circ$ ROIs described in \S~\ref{sec:CountingAnalysis}.  We note that the \Fermi~Science Tools do not currently include a means of performing a likelihood analysis on data selected using an energy-dependent ROI, as is possible with the counting analysis.

 \subsection{Background Selection} \label{sec:BackgroundSelection}

The proper selection of background regions with which to compare the stacked signal is a crucial component to the counting analysis and provides a control sample for the joint likelihood analysis.  Since the spacecraft's geomagnetic coordinates and its celestial pointing can vary significantly over the 2700 s duration under consideration, the use of an off-and-on source method of background determination may not always adequately represent the true background.  In addition, the use of intervals immediately prior to the burst trigger or after the end of the prompt emission, as observed by GBM or BAT, to estimate a background interval could serve to bias our investigation, as LAT-detected GRBs have exhibited both delayed and extended high-energy emission on timescales that exceed the durations traditionally defined by observations in the keV$-$MeV energy range \citep{Ackermann2013} and subthreshold emission prior to the burst trigger cannot be ruled out. Therefore, we avoid using intervals immediately preceding or succeeding the burst activity for background estimation.  

Instead, we attempt to locate an interval at least one orbit prior to the GRB trigger that best matches the spacecraft's observing conditions at the time of the trigger.  This includes matching the same off-axis angle between the GRB sky coordinates and the LAT boresight, the spacecraft's geomagnetic coordinates in orbit, and the angle of the GRB location to the Earth's zenith at the time of trigger. These three criteria serve to match the charged-particle background, the Galactic and isotropic backgrounds, and possible contamination from Earth limb photons, respectively, to that observed by the LAT at the time of the GRB trigger.  A period of 30 sidereal orbits (171,915 s) prior to trigger adequately matches these geomagnetic and pointing criteria and provides an interval with which to estimate the background during the GRB. An example of the spacecraft's orbit and orientation with respect to the position of a GRB in our sample and its associated background orbit can be seen in Figure \ref{Fig:OrbitalComparison}. 

We note that this selection does not adequately represent the expected background for the 15 GRBs that either triggered ARRs of the spacecraft or occurred while the spacecraft was performing a ToO, since maneuvers take the spacecraft out of survey mode and initiate a custom pointing that keeps the GRB in the LAT FOV.  As a result, there is no previous interval that matches the spacecraft's orbit and orientation during the repoint, and as such, these bursts are excluded from the counting analysis.  Since the background for each individual GRB is being modeled in the likelihood analysis, this method can adequately take into account the spacecraft motion for these bursts.  This leaves a total of 64 bursts for which we can apply the counting analysis.  Therefore, the application of the likelihood technique to the bursts for which we have the most comprehensive observations is a significant advantage of this method over the counting analysis. 

In order to validate the effectiveness of this background selection for bursts that did not trigger an ARR or occurred during a ToO, we extract the observed photons over a 2700 s interval from 1000 random locations on the sky covering an energy range from 75 MeV to 300 GeV, using a 10$^\circ$, 12$^\circ$, and energy-dependent ROI.  We then extract the observed photons at the same location, but 30 sidereal orbits prior to the original selection interval.  We can then test whether the observed photons during the signal and background intervals are consistent with being drawn from the same distribution by examining the resulting significance distribution.  We define the counting significance as $(S-B)/\sqrt{S+B}$, where $S$ and $B$ are the counts in the signal and background intervals, respectively \citep{LiMa83}.

The resulting distribution for the 12$^\circ$ ROI has a mean value of $\mu_{12} = 0.09$ and variance of $\sigma_{12} = 1.01$.  Among our 1000 trials, we obtain three false positives above $3\sigma$, consistent with expectations, as we expect 99.7$\%$ of values drawn from our significance distribution to lie within $3\sigma$ of the mean.  The results for the 10$^\circ$ and energy-dependent ROIs were consistent with these values.  Therefore, our background selection is found to be robust for any single source and background interval selection.

\section{Results} 
\label{sec:Results}

\subsection{Counting Analysis} \label{sec:ResultsCountingAnalysis}

\begin{figure}[t]
\includegraphics[width=1\columnwidth]{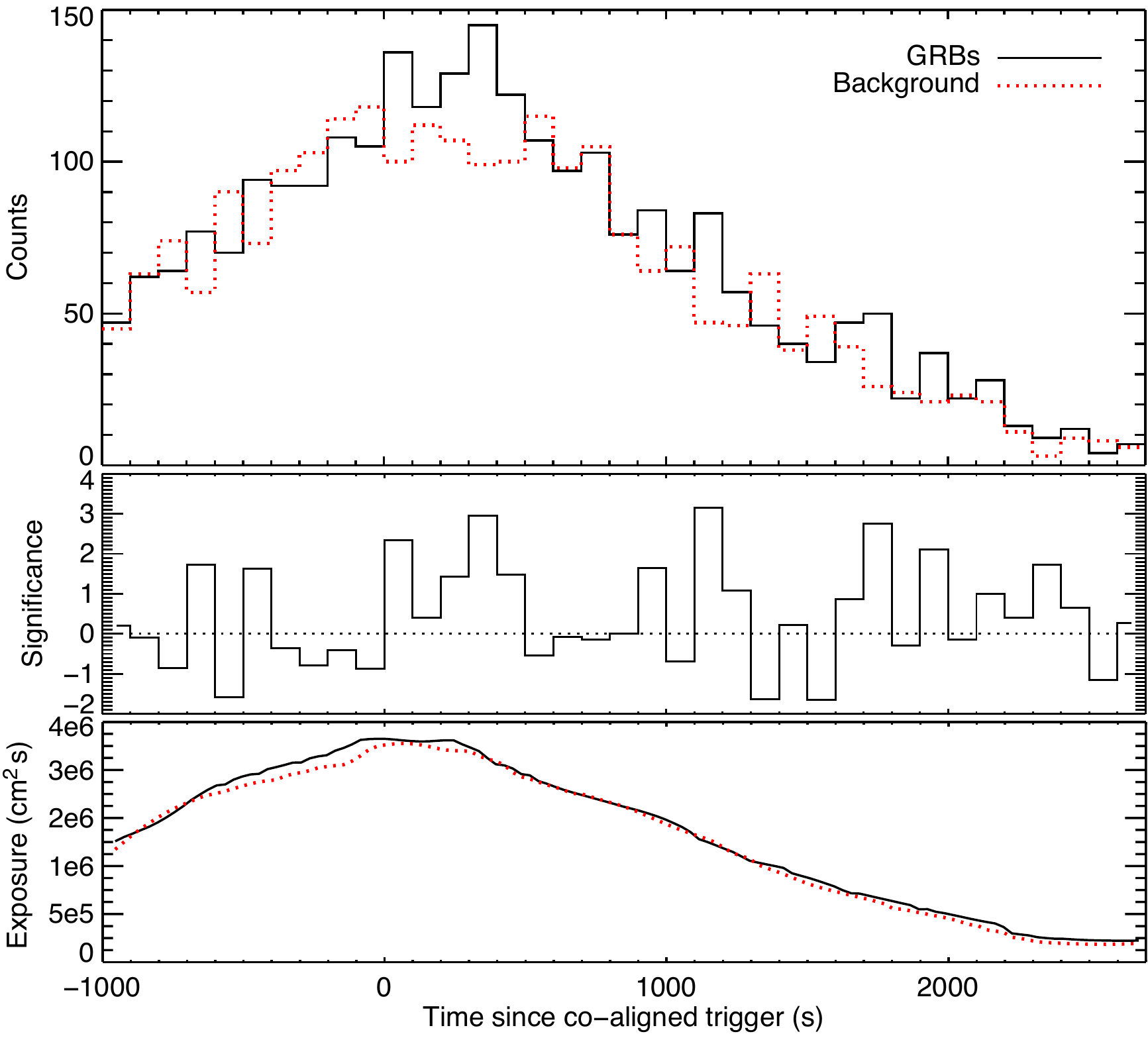}

\caption{Top panel: a stacked light curve of signal (black solid line) and background (red dotted line) counts for a 10$^\circ$ ROI selection, binned to 100 s intervals. Middle panel: The significance of the signal compared to the background. Bottom panel:  comparison of the total exposure during the signal and background orbits.}
\label{Fig:StackedLightCurve-ARRs}
\end{figure}

Using the data selection criteria described in Section~\ref{sec:CountingAnalysis}, we can directly compare the extracted signal to the counts accumulated using the same selection criteria, but collected during the background orbits described in Section~\ref{sec:BackgroundSelection}.  When considering the entire interval from $T_{0}$ to 2700 s and an energy range from 75 MeV to 30 GeV, we obtain a total accumulated signal and background of $S_{10} = 1711$ and $B_{10} = 1476$, $S_{12} = 2365$ and $B_{12} = 2161$, and $S_{\rm EROI} = 750$ and $B_{\rm EROI} = 626$ counts for a significance of $4.16 \sigma$, $3.03 \sigma$, and $3.34 \sigma$ for the 10$^\circ$, 12$^\circ$, and energy-dependent ROIs, respectively.  
 
Introducing cuts on the data, as we have here with the extraction radius, in order to maximize the observed significance introduces a bias in the analysis due to the so-called look-elsewhere effect.  The chance that the observed significance could have arisen at random owing to the size of the parameter space that was searched can be accounted for by applying trial factor corrections to the final significance.  Here we searched the data using three different extraction radii, and therefore the final significance needs to be attenuated as $1-(1-\alpha$)$^N$, where $\alpha$ is the probability of observing such a value by chance\footnote{Recall that for a normal distribution, a $3\sigma$ detection has a probability of chance occurrence of $\alpha = 1-99.7\% = 0.003$} of the detection and $N$ represents the number of trials.  For the analysis presented above, our $4.16 \sigma$ detection therefore becomes an erf($\frac{4.16}{\sqrt{2}})^{3}$ $\sim 3.9 \sigma$ detection.  We note that this correction factor assumes that each data set is statistically independent, which is overly conservative in our case, since the three different ROI selections result in data sets that are subsets of each other.  Therefore, we regard the $3.9 \sigma$ detection significance as a conservative lower limit to the true significance of the signal over the background.

Focusing on the 10$^\circ$ ROI analysis, we can create a stacked light curve that is co-aligned to the trigger time of each GRB.  This light curve of stacked signal and background counts, binned to 100 s intervals, is shown in the top panel of Figure \ref{Fig:StackedLightCurve-ARRs}, with the middle panel showing the Gaussian significance of their difference and the bottom panel showing the summed LAT effective area during the observations.  The overall shape of the light curve reflects the evolution in the total effective area.  The observed count rate rises as additional fields containing GRBs enter the LAT FOV and falls as they exit or as the spacecraft enters the (SAA), whereby data taking is disabled.  The total effective area, and hence the stacked counts light curve, peaks near the co-aligned trigger, as all fields are predicated to be in the LAT FOV at this time owing to our sample selection criteria.  The stacked light curve shows a $\sim3\sigma$ excess roughly 30 s after the co-aligned trigger, followed by additional periods of excess signal over background, although a consideration of trial factors lessens the significance of any one peak in such a time-resolved analysis.  

\begin{figure}[t]
\includegraphics[width=1\columnwidth]{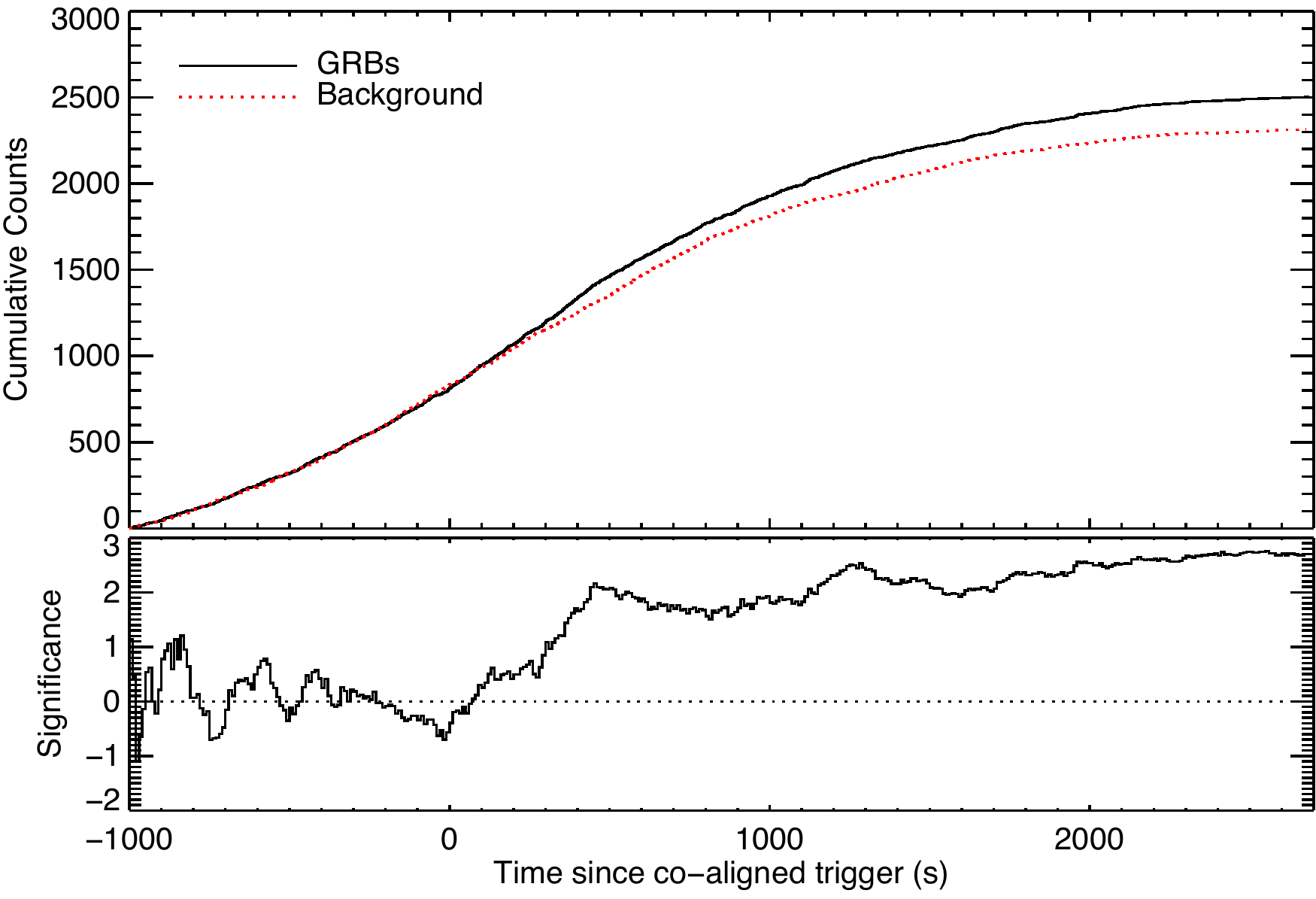}
\caption{Top panel: cumulative signal (black solid line) and background (red dotted line) counts as a function of time. Bottom panel: significance of the cumulative signal compared to the cumulative background for a 10$^\circ$ ROI selection.}
\label{Fig:CumulativeLightCurve-ARRs}
\end{figure}

The median ratio of the exposure during the signal and background orbit is found to be $\mu = 1.05$, with a variance of $\sigma = 0.072$, highlighting that the exposures of the signal and background orbits are well matched, but not exactly equal. Small mismatches in exposure such as these may be a fundamental limitation of the stacked counting analysis performed here.  Such differences are small for the source/background comparison for any one burst, but when adding 79 such comparisons, these small mismatches contribute to a non-negligible difference in the total summed exposure, which can be seen in the bottom panel of Figure \ref{Fig:StackedLightCurve-ARRs}. Ultimately, normalizing the observed counts by the estimated exposure for each burst and performing a summed rate comparison can account for these differences.

Next, we examine the difference between the cumulative signal and background over much wider time intervals.  In the top panel of Figure \ref{Fig:CumulativeLightCurve-ARRs}, we show the cumulative signal and background counts as a function of time since $T_{0}$ - 1000 s for a 10$^\circ$ ROI selection, with the bottom panel again showing the significance of their difference.  The effective area of the stacked observations drops to zero above $T_{0}$ + 2700 s as all fields exit the LAT FOV, and as such, the cumulative light curve levels off accordingly.  The excesses seen in Figure \ref{Fig:StackedLightCurve-ARRs} can be clearly traced as local maxima in Figure \ref{Fig:CumulativeLightCurve-ARRs}. Although the difference between the signal and background varies above and below zero significance prior to the co-aligned trigger, the signal clearly begins to diverge from the background at $T_{0}$, above which the significance climbs to approximately $\sim3 \sigma$ at $T_{0}$ + 2700 s. We note that this cumulative significance  differs from that quoted above because the integration period here begins at $T_{0}$ - 1000 s.

\begin{figure}[t]
\includegraphics[width=1\columnwidth]{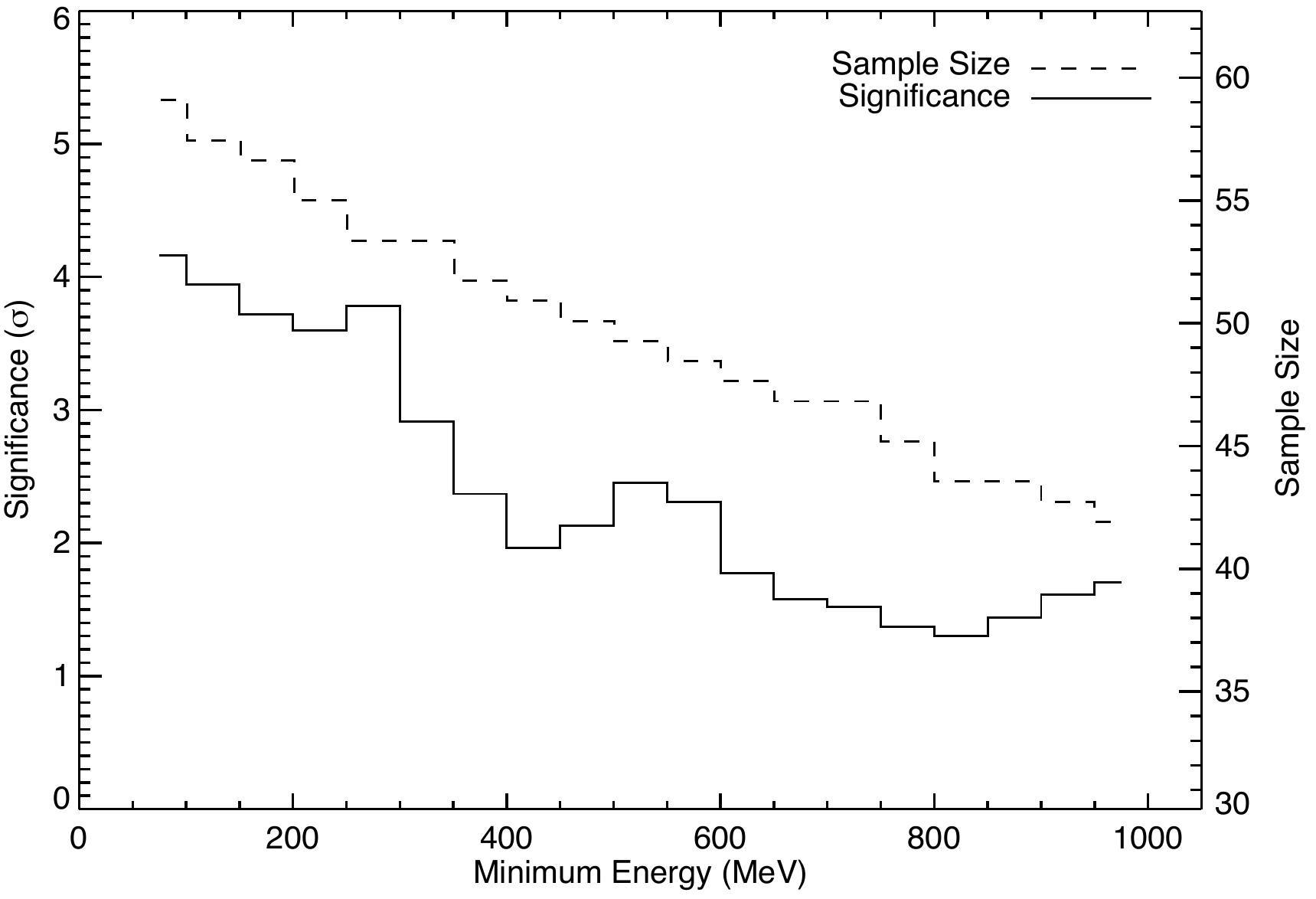}
\caption{Counting analysis: the dependence of the signal significance (solid line) and the sample size (dashed line) on the minimum selection energy $E_{\rm min}$ for a 10$^\circ$ ROI.}
\label{Fig:SignificancevsEnergy}
\end{figure}

We also examine the dependence of the signal significance on the minimum energy $E_{\rm min}$ used in our selection criteria. In Figure~\ref{Fig:SignificancevsEnergy}, we plot the resulting signal significance as a function of $E_{\rm min}$ when considering the interval from $T_{0}$ to 2700 s for a 10$^\circ$ ROI selection, showing that the detection significance does not improve when only considering higher-energy photons.  The dashed line, representing the number of bursts with photons contributing to the modified selection criteria, falls steadily with increasing $E_{\rm min}$.  
 
Finally, an identical analysis of the preburst observations, covering an interval from $T_{0}$ - 2700 s to the co-aligned trigger, reveals no excess emission above the background.  For this control sample, using a 10$^\circ$ ROI selection, we obtain a total accumulated signal and background of $S = 1704$ and $B = 1681$ counts, respectively, for a significance of $0.40 \sigma$.  

\subsection{Joint Likelihood Analysis} \label{sec:JointLikelihoodAnalysis}

Using the data selection criteria described in Section~\ref{sec:CountingAnalysis} and the analysis method outlined in Section~\ref{sec:CompositeLikelihoodAnalysis}, we first performed a joint likelihood analysis on the ensemble of GRB locations, including those that triggered ARRs of the spacecraft or occurred during ToOs, for an integration period covering the entire 2700 s post trigger and an energy range from 75 MeV to 30 GeV, using both a 10$^\circ$ and 12$^\circ$ ROI.  The resulting TS of the joint likelihood fit is 59 and 58.6, respectively, roughly representing a $\sim7 \sigma$ detection significance of a point source being present in excess to the expected background using either ROI size.  The TS of an identical analysis performed on the background orbits defined in Section~\ref{sec:BackgroundSelection} is consistent with zero. 

\begin{figure}[t]
\includegraphics[width=1\columnwidth]{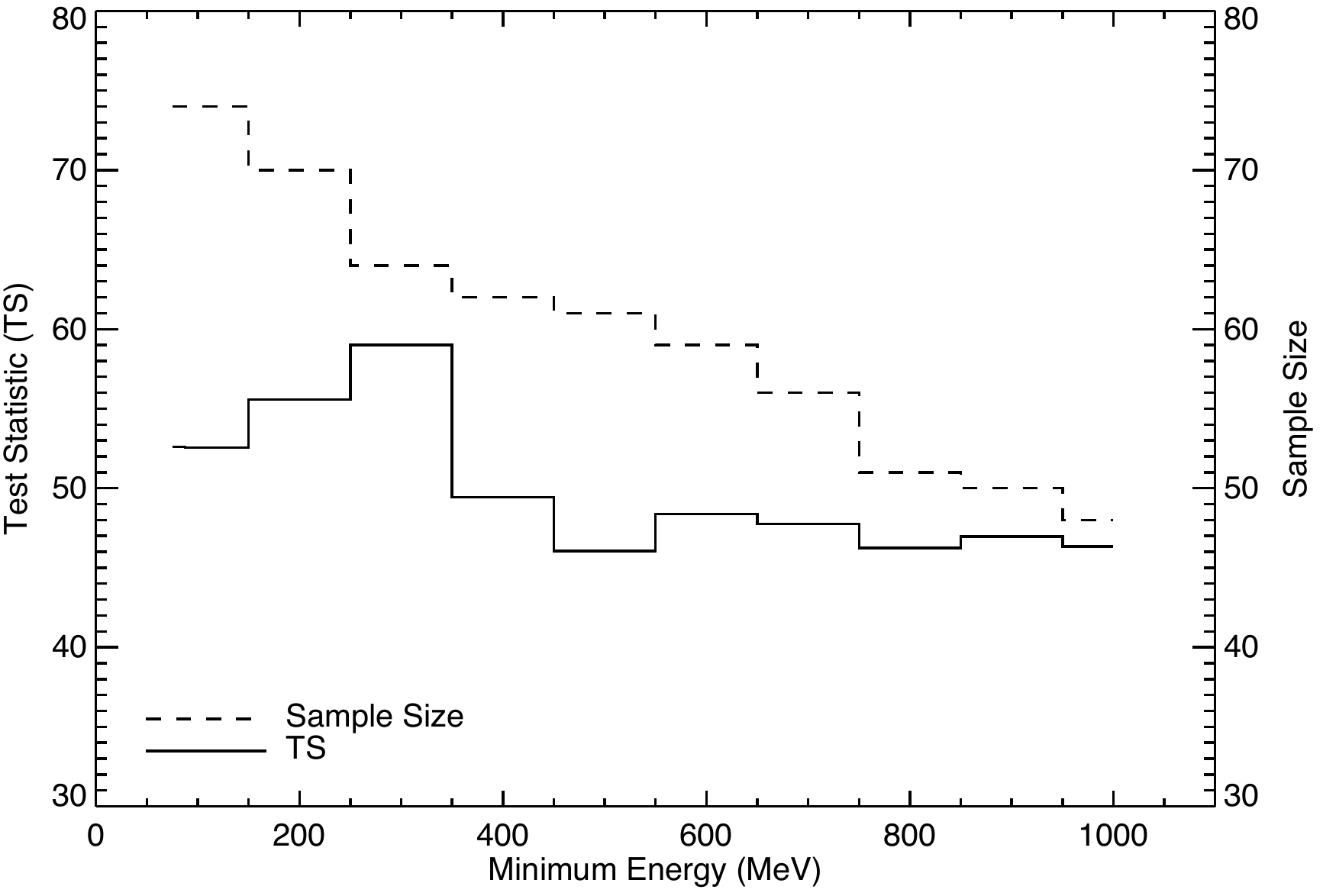}
\caption{Likelihood analysis: the dependence of the likelihood TS (solid line) and the sample size (dashed line) on the minimum selection energy $E_{\rm min}$ for a 10$^\circ$ ROI..}
\label{Fig:TSvsEnergy}
\end{figure}

Focusing on the 10$^\circ$ ROI analysis, we examine the dependence of the TS value on the minimum energy $E_{\rm min}$ used in our selection criteria. In Figure~\ref{Fig:TSvsEnergy}, we plot the resulting TS for a range of $E_{\rm min}$ values, showing that the detection significance rises as low-energy photons are excluded from the fit, peaking at $E_{\rm min} = 300$ MeV, before eventually leveling out above 400 MeV.  The dashed line represents the number of analysis intervals for which we observed photons matching the selection criteria.  As $E_{\rm min}$ rises, the number of intervals with photons contributing to the joint likelihood analysis falls, with nearly one-third of the burst positions being removed from the sample when $E_{\rm min} = 1000$ MeV.  Because the detection significance peaks at $E_{\rm min} = 300$ MeV, we will focus the remainder of our joint likelihood analysis on an energy selection criterion covering 300 MeV to 30 GeV. 
 
\begin{figure}[t]
\includegraphics[width=1\columnwidth]{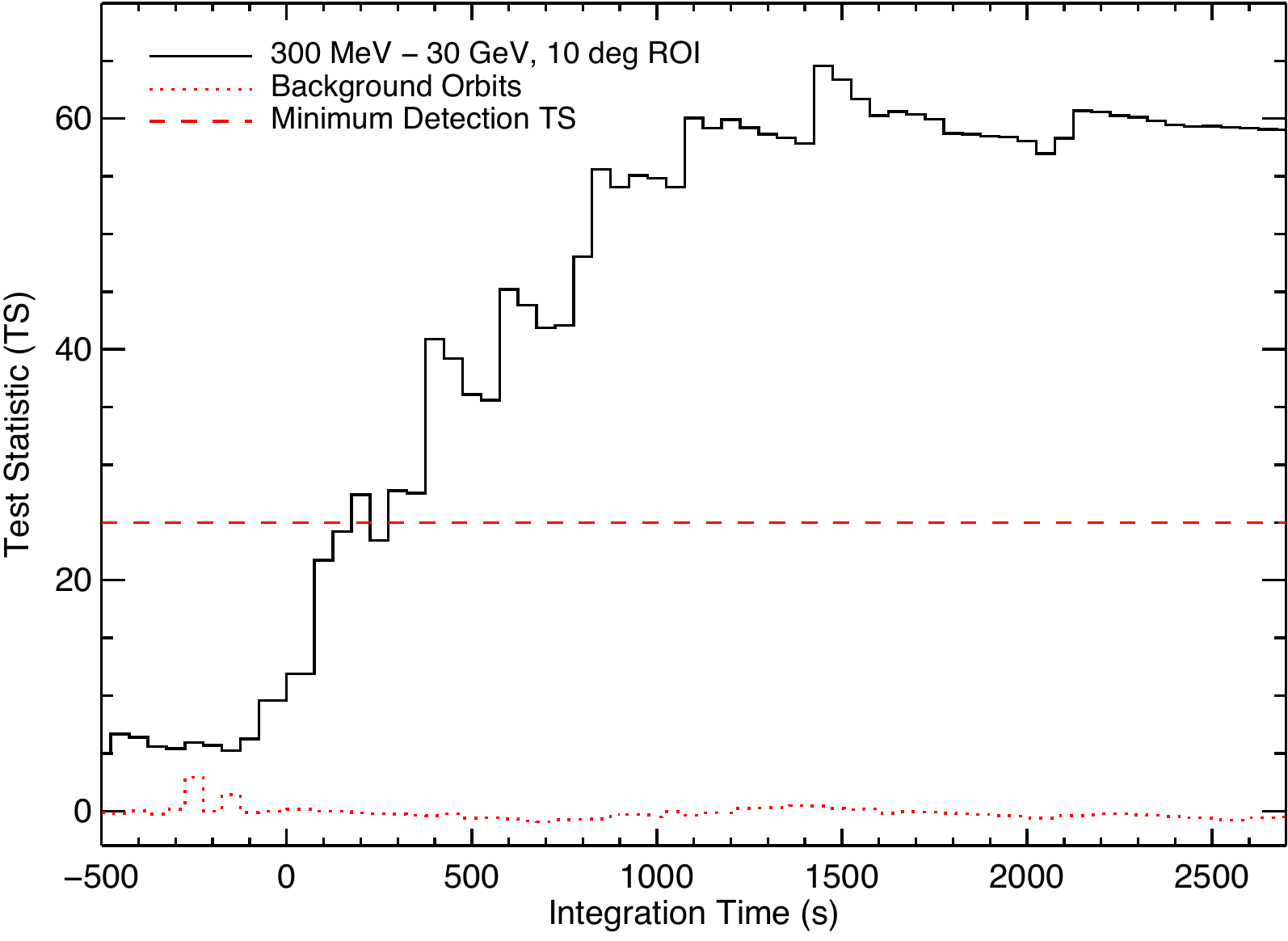}
\caption{Likelihood TS (black solid line) as a function of integration time from trigger, binned to 100 s intervals, for the entire GRB sample.  The red dotted line represents the same analysis applied to designated background intervals exactly 30 sidereal orbits prior to trigger. The red dashed line represents the minimum TS that we assign to constitute a detection.}
\label{Fig:MaxTSVsTime}
\end{figure}

In order to investigate the time dependence of the excess signal, we also performed a joint likelihood analysis over a range of integration times before and after the trigger time, going from from $T_{0}$ - 1000 s to $T_{0}$ and from $T_{0}$ to $T_{0}$ + 2700 s, in 50 s intervals..  The TS as a function of integration time is shown in Figure~\ref{Fig:MaxTSVsTime}.  The red dotted line represents the TS of an identical analysis performed on each burst's associated background orbits. The TS of the joint likelihood fit to the data collected over the 1000 s prior to the co-aligned trigger is consistent with zero.  The TS rises slightly for shorter integration duration approaching $T_{0}$, but is below TS = 9, roughly equivalent to $3\sigma$, for all integration durations prior to $T_{0}$, except for the 50 second interval covering $T_{0}$ - 50 to $T_{0}$, at which point it approaches TS $\sim10$.  At all subsequent integration times after $T_{0}$, the significance of the signal excess above background is greater than $3\sigma$, reaching $5\sigma$ within 200 s after $T_{0}$.  The TS values exhibit local maxima as a function of time, reflecting the arrival of photons in excess to our background model, before leveling off at TS $= 59$ as the accumulated effective area drops to zero.  The data collected from the same fields during the background intervals show no such excess when fit to our background models, being consistent with TS $\sim$ 0 for roughly all integration periods under consideration. 

Before we can assess the final significance of our likelihood analysis, we again need to take trial factors into account.  Since we considered two different ROI sizes and 10 different minimum energy boundaries, this introduces $N = 12$ trials.  Therefore, a TS $= 59$, or $\sqrt{59} \sim 7.68\sigma$, detection becomes an erf($\frac{7.68}{\sqrt{2}})^{12}$ $\sim 7.46 \sigma$ detection.  Again, since the ROI and minimum boundary selections did not result in statistically independent data sets, we regard this attenuated significance as a conservative lower limit to the true significance of the detection. 

\begin{figure}[t]
\includegraphics[width=1\columnwidth]{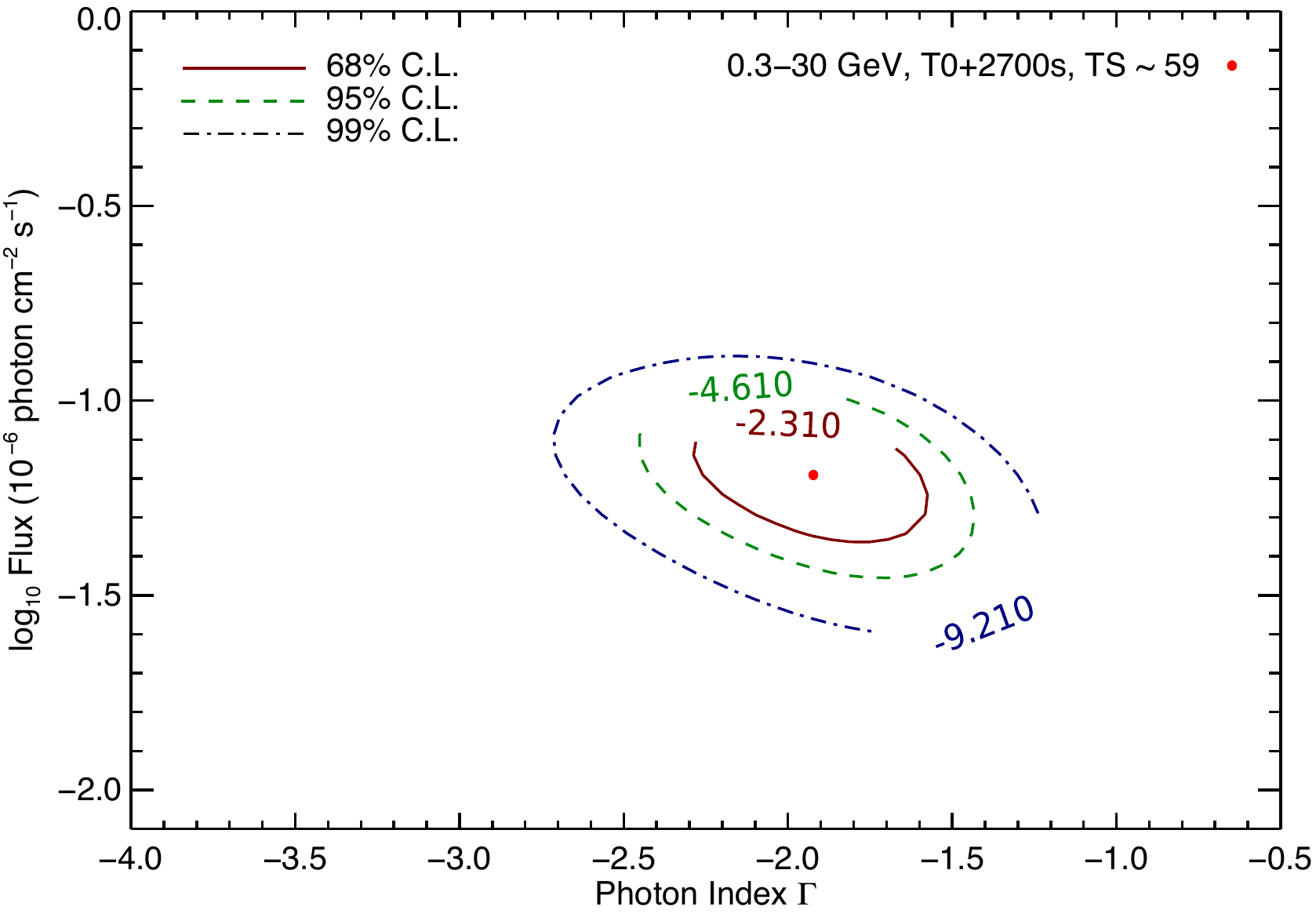}
\caption{Contours of the joint likelihood profile as a function of photon flux and photon index, generated from data covering 2700 s post trigger and a 300 MeV to 30 GeV energy range.}
\label{Fig:LikelihoodContourPlot}
\end{figure}

By scanning the joint likelihood function over a range of photon indices and flux normalizations for the point source in our model, we can obtain a joint likelihood profile that is a function of these two common parameters of interest.  Finding the maximum of this profile allows us to estimate the stacked source flux and characteristic photon index of the sample.  A contour plot of the resulting joint likelihood profile for the analysis covering the entire 2700 s interval post trigger and an energy range from 300 MeV to 30 GeV is shown in Figure \ref{Fig:LikelihoodContourPlot}.  The best-fit photon flux and photon index of the combined data is $F_{\rm ph} = 6.4\times10^{-8}$ photons cm$^{-2}$ s$^{-1}$ and $\Gamma =  -1.92$, respectively.  The 68$\%$, 95$\%$, and 99$\%$ confidence levels (C.L.) for the photon flux estimate are shown as the solid, dashed, and dash-dotted lines, respectively.

For comparison, a typical photon flux upper limit for an individual well-observed GRB that remained in the LAT FOV for an entire 2700 s duration is on the order of $F_{\rm ph,UL} \sim 10^{-6}$ \citep{Ackermann2012}, making it clear why none of the bursts in our sample were individually detected.  The best-fit photon index is representative of indices measured at late times in previously detected LAT bursts (e.g. GRB~110731A) and consistent with photon indices measured by XRT of GRB afterglows at late times. 

\begin{figure}
\includegraphics[width=1\columnwidth]{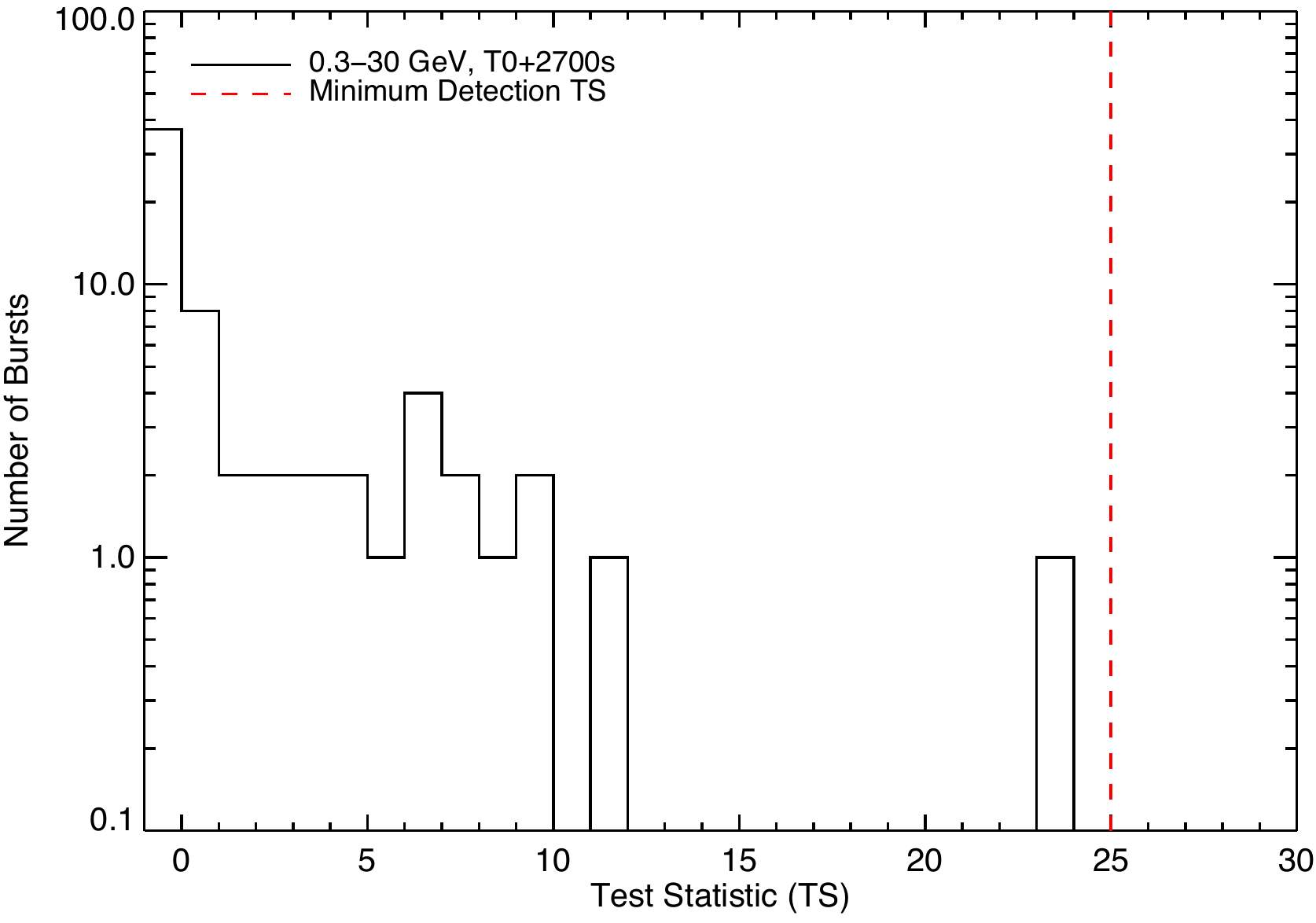}
\caption{Distribution of individual TS values for a likelihood analysis covering the 2700 s after co-aligned trigger.  The red dashed line represents the minimum TS that we assign to constitute a detection. }
\label{Fig:TSDistribution_GRBs_300Mevto30GeV_-10sto2700s}
\end{figure}

In order to ensure that one or more GRBs are not dominating the observed excess seen in Figure~\ref{Fig:MaxTSVsTime}, we examine the TS distribution for the 2700 s second integration period succeeding $T_{0}$ in Figure~\ref{Fig:TSDistribution_GRBs_300Mevto30GeV_-10sto2700s}.  The distribution peaks at TS $= 0$ for 41 GRBs, nearly half the sample under consideration.  The remaining bursts are evenly distributed between $1 \lesssim$ TS $\lesssim 10$, with one burst (GRB~110903A) at TS $= 23$.  This burst is just under the fiducial $5\sigma$ threshold for the a burst to be considered detected by the LAT.  Removing this burst from the sample, we find that the TS of the joint likelihood analysis for the 2700 s second integration interval drops from TS = 55 to TS = 41, still yielding a strong detection of the remaining bursts. Moreover, the best-fit flux and index remain relatively unchanged, at $F_{\rm ph} = 6.0\times10^{-8}$ photons cm$^{-2}$ s$^{-1}$ and $\Gamma =  -2.09$.

\subsection{Comparison of the Two Methods} \label{sec:MethodComparison}

\begin{figure}
\includegraphics[width=1\columnwidth]{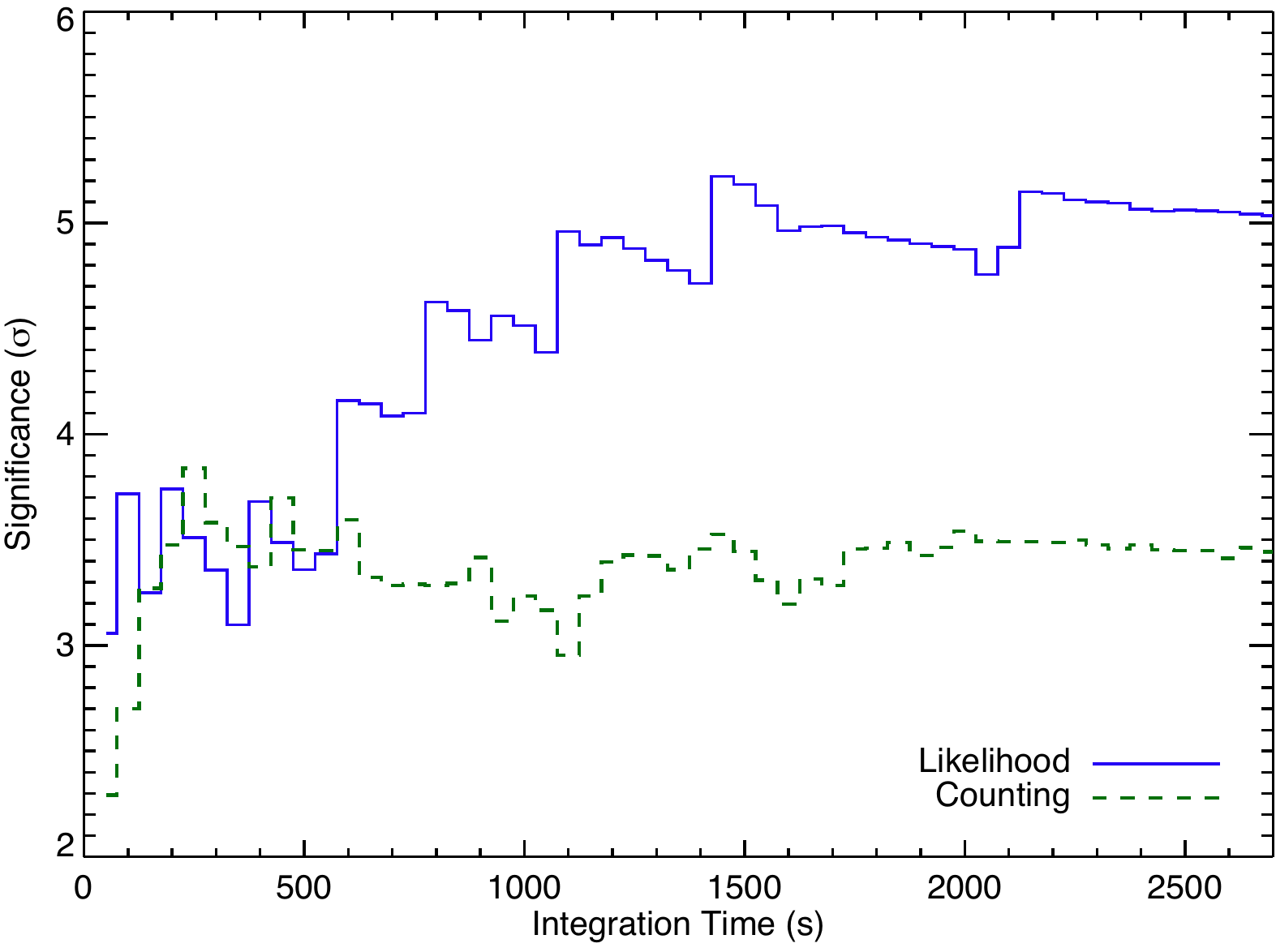}
\caption{Comparison of the significance of the detected signal as a function of integration time, covering an energy range from 300 MeV to 30 GeV for a 10$^\circ$ ROI selection, for both the counting and likelihood techniques.}
\label{Fig:LikelihoodVsCounting_v2}
\end{figure}

The counting and joint likelihood analyses employed here have both revealed the existence of a signal in excess of the expected background for the GRB locations in our sample.  Here we compare the two methods to determine whether there are any inherent benefits to using one method over the other.  The significance of the detected signal as a function of integration times, covering an energy range from 300 MeV to 30 GeV using a 10$^\circ$ ROI, for each method is displayed in Figure \ref{Fig:LikelihoodVsCounting_v2}.  In order to ensure a proper comparison, we have removed the bursts that triggered ARRs of the spacecraft or occurred during ToOs from the likelihood analysis. 

With the ARR and ToO bursts removed, the joint likelihood method provides stronger evidence for a signal excess when considering the entire 2700 s integration period, yielding a total signal significance of $\sim5 \sigma$, whereas the counting technique provides only a marginal detection at $\sim3.4 \sigma$.  On smaller integration time, the two methods are roughly consistent, before diverging for longer integration intervals.  The significance derived from the joint likelihood analysis rises gradually as more data are accumulated before leveling out to roughly $5 \sigma$, whereas the significance derived from the counting analysis remains largely unchanged through the inclusion of these data.  Note that the final signal significance quoted here for the counting analysis differs from that reported in \S\ref{sec:ResultsCountingAnalysis} because of the different energy ranges under consideration.  Here we are examining photons over a 300 MeV to 30 GeV energy range, in order to match the analysis performed using the joint likelihood technique. 

\subsection{Population Demographics} \label{sec:PopulationDemographics}

\begin{figure}
\includegraphics[width=1\columnwidth]{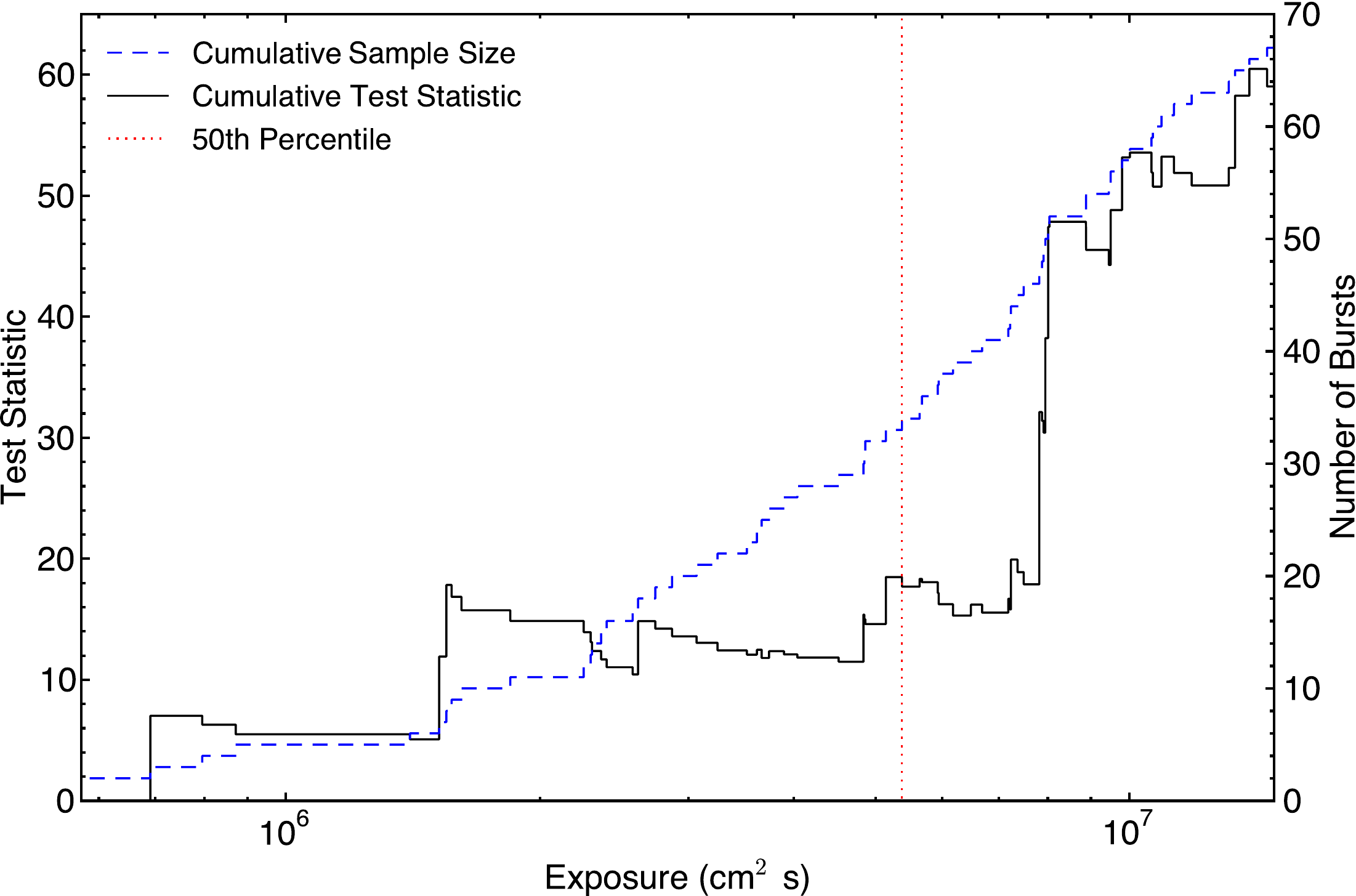}
\caption{Cumulative likelihood TS as a function of burst exposure.}
\label{Fig:TSVsExposure}
\end{figure}

In order to examine which bursts in our sample contribute most significantly to the observed signal, we calculate the cumulative TS by rerunning the likelihood analysis for each additional burst, sorted as a function of a selected burst property.  An example of this method can be seen in Figure \ref{Fig:TSVsExposure}, where the cumulative TS is plotted as a function of burst exposure.  Here exposure is defined as an integral of the total response over the entire ROI\footnote{http://fermi.gsfc.nasa.gov/ssc/data/analysis/documentation/Cicerone}.  The TS is first calculated for the burst with the lowest exposure, covering an energy range from 300 MeV to 30 GeV and a period of 2700 s post trigger, then recalculated by including the data from the burst with the next highest exposure, and so on.  The result is an increasing cumulative TS that eventually peaks at the value reported in Section~\ref{sec:JointLikelihoodAnalysis}.  The dashed line in Figure \ref{Fig:TSVsExposure} represents the cumulative sample size as a function of exposure, and the dotted line demarcates the 50th percentile of the sample.  If all the bursts contributed equally to the final TS value, independent of their exposure, then one would expect the cumulative TS to rise at the same rate as the cumulative sample size.  This is not the case in Figure \ref{Fig:TSVsExposure}, where the bursts with exposures in the first 50th percentile of the sample contribute very little to the final signal significance.  This would be expected, since longer observations of bursts closer to the instrument boresight would be more sensitive at detecting extended subthreshold emission. Still, Figure \ref{Fig:TSVsExposure} shows that exposure alone is not entirely indicative of whether a burst will contribute to the final TS value.  Several bursts with the highest exposure are seen to contribute very little to the overall signal, whereas several of the bursts with the lowest exposure do contribute to the significance of the final signal. This is also reflected by the fact that the non-ARR sample of bursts discussed in Section~\S\ref{sec:MethodComparison} is still detected at roughly $\sim5 \sigma$ (TS$\sim26$), compared to $\sim7.7\sigma$ (TS$\sim59$) for the entire sample.  Therefore, the ARR and ToO bursts contribute most, but not all, of the observed signal.

\begin{figure}
\includegraphics[width=1\columnwidth]{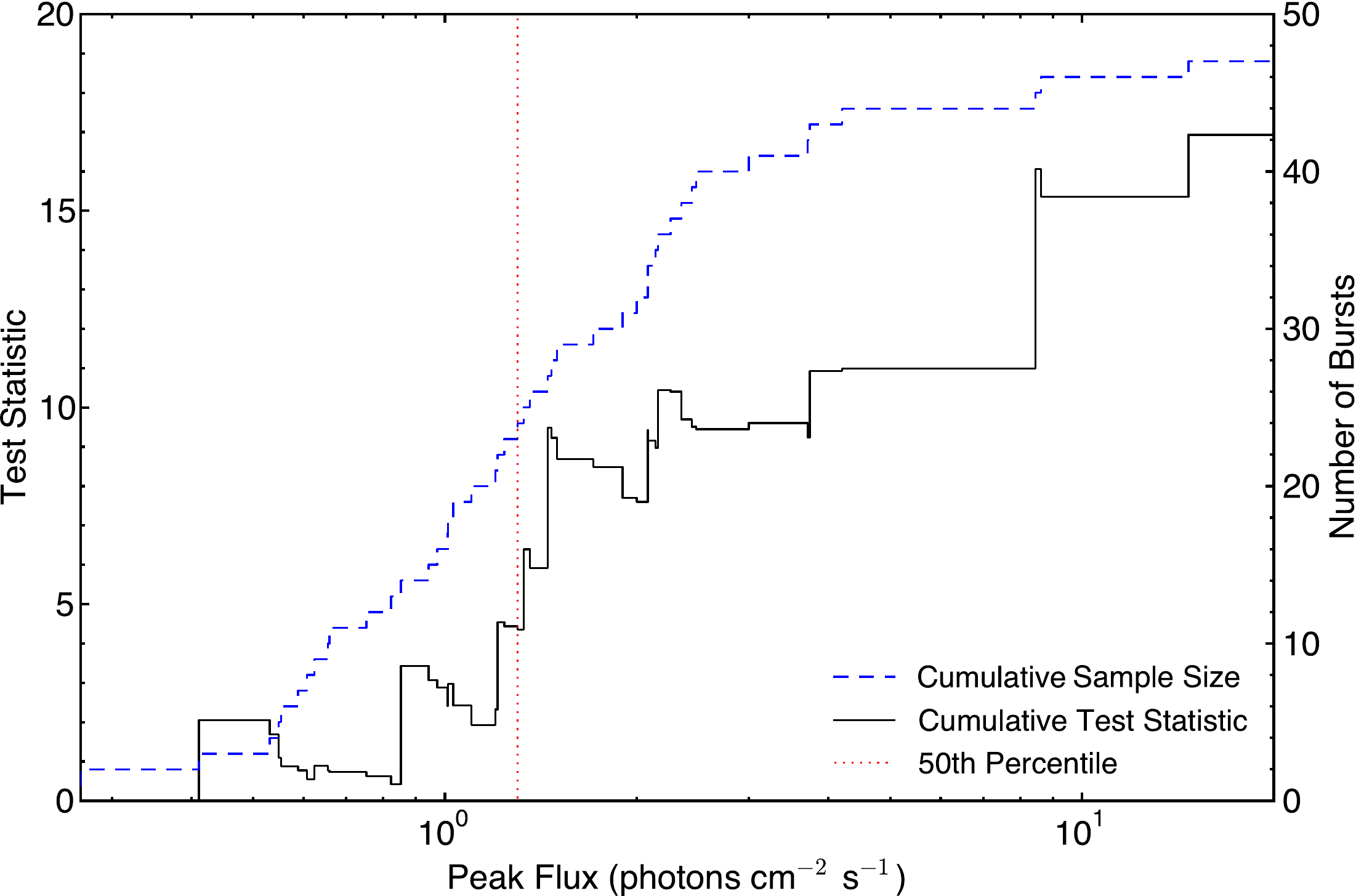}
\caption{Cumulative likelihood TS as a function of peak flux as measured by BAT over a 15-150 keV energy range.}
\label{Fig:TSVsPeakFlux}
\end{figure}

We can test whether the observed low-energy gamma-ray flux of a source plays an additional role in its detectability and hence contribution to the final TS value.  To examine this, we calculated the cumulative TS as a function of the burst's peak photon flux in the 15-150 keV energy range as measured by BAT \citep{Donato2012}, and we show the results in Figure \ref{Fig:TSVsPeakFlux}.  The bursts with the highest observed peak flux contribute strongly to the cumulative TS, although there remain bursts with flux values above 1.0 photon cm$^{-2}$ s$^{-1}$ that do not appear to contribute significantly to the final TS value.  Note that the sample size presented in Figure \ref{Fig:TSVsExposure} differs from that in Figure \ref{Fig:TSVsPeakFlux} because not every burst in the likelihood sample was initially detected by BAT, and therefore a photon flux estimate of their prompt emission is not available. 

Given the extended nature of the signal inferred in Figure \ref{Fig:LikelihoodVsCounting_v2}, we also calculate the cumulative TS as a function of the burst's X-ray brightness at 11 hrs, in the 0.3-10 keV energy range, as measured by XRT.  The X-ray flux at 11 hr has become a standard measure of afterglow brightness, as it samples the afterglow light curve at a period where the steep decay and plateau phases have typically ceased \citep{Nousek06}.  To calculate these values, we downloaded the XRT data for each from the \Swift~XRT light-curve repository \citep{Evans09} and fit the flux light curves with a multi-segmented afterglow model defined by \citet{Zhang06}.  The fitting procedure employed is outlined in \citet{Racusin09}.  The result of the X-ray brightness versus cumulative TS analysis is presented in Figure \ref{Fig:TSVsXRTFlux_11hrs}.  There exists an even stronger trend of bursts with bright X-ray emission 11 hrs post trigger contributing significantly to the final TS value, compared to the prompt gamma-ray flux measured by the BAT.  In fact, fewer than half of the brightest bursts in the sample contribute most of the signal, with the addition of the remaining bursts actually decreasing the final signal significance.

\section{Discussion} \label{sec:Discussion}

The stacking analysis presented above shows significant evidence for subthreshold emission among LAT non-detected GRBs.  These results are consistent with the conclusions drawn from the first \Fermi~LAT GRB catalog \citep{Ackermann2013}, in which the authors find a significance correlation between the burst fluence as measured by GBM in the 10 keV$--$1 MeV energy range, the LAT boresight angle at the time of trigger, and burst detectability (see their Figure 31).  The authors find that LAT-detected GRBs are among the most fluent bursts observed by the GBM, but that this fluence threshold for a LAT detection falls with decreasing boresight angle, tracing the instrument sensitivity as a function of off-axis angle.  

The counting and likelihood analysis techniques employed above show evidence for subthreshold emission on both prompt and extended timescales, mirroring the range of emission timescales observed in the LAT-detected population. The observed signal detected by both techniques remains significant over the entire 2700 s period under consideration and suggests that the extended emission observed in some LAT-detected GRBs may be common among the population.  The photon index measured through the likelihood analysis is consistent with the average value measured for the LAT-detected population \citep{Ackermann2013}.  This value is also consistent with the photon index expected from the high-energy extension of the synchrotron spectrum due to the external forward shock in the standard afterglow theory, supporting an external shock origin of the extended emission.

Our detection of ubiquitous long-lived emission, albeit below the LAT detection threshold, is consistent with similar work performed by \citet{Lange13}.  In their work, the authors examined a sample of 99 GBM localized GRBs that were not detected by LAT and found evidence for long-lived subthreshold emission that lasted for as much as 10 times the bursts' $T_{\rm 90}$ duration in the keV-MeV energy range. Unlike the conclusions drawn by \citet{Lange13}, though, we find that the photon index of the subthreshold population derived from our joint likelihood analysis is largely consistent with that observed in the LAT-detected bursts.  The different results may be due to the different techniques employed, with the hardest bursts in our sample contributing most significantly to the joint likelihood results, and hence possibly providing a harder average photon index than that measured by \citet{Lange13}.

\begin{figure}
\includegraphics[width=1\columnwidth]{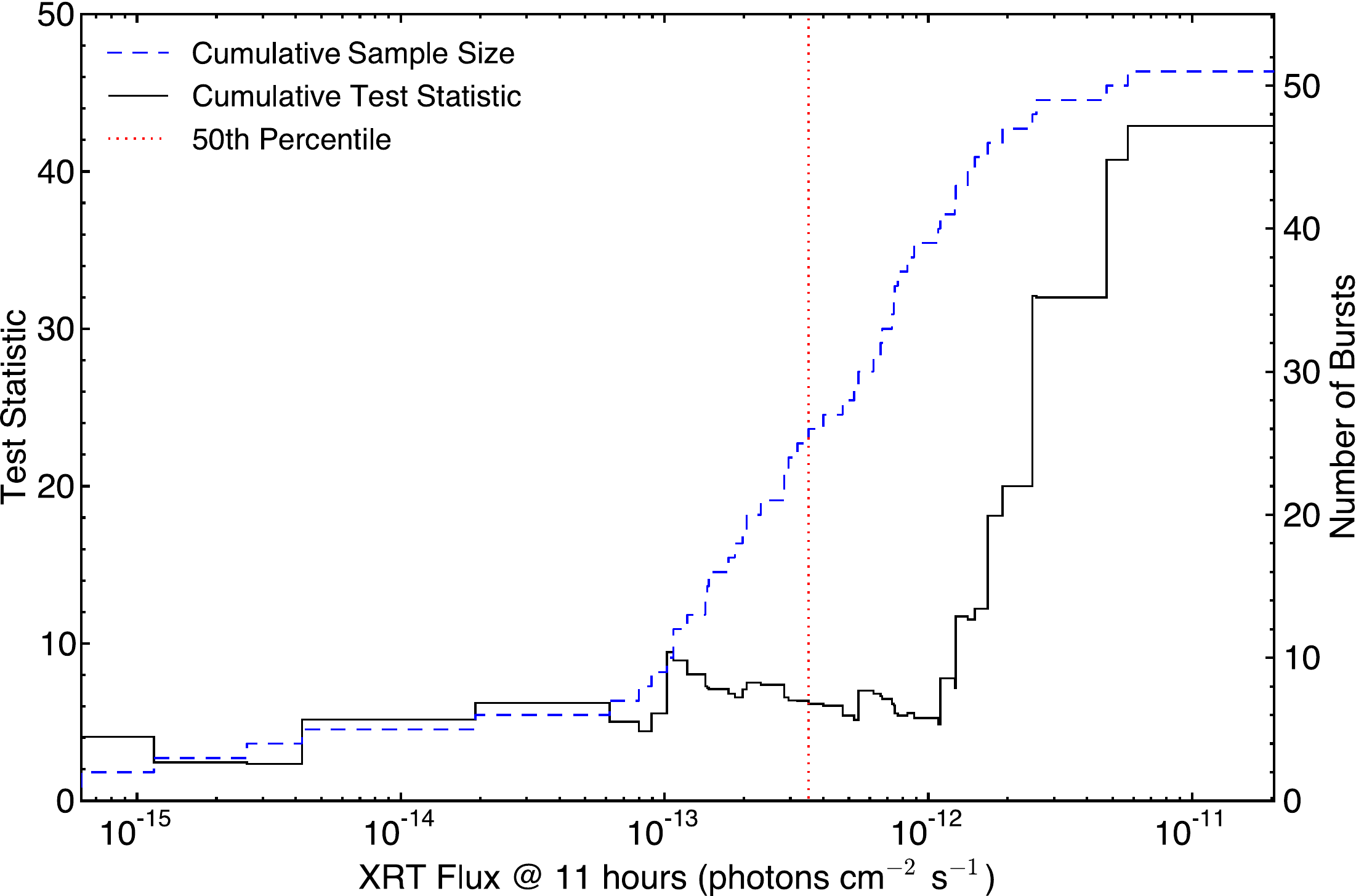}
\caption{Cumulative likelihood TS as a function of X-ray flux as measured at 11 hrs post trigger by XRT over a 0.3-10 keV energy range.}
\label{Fig:TSVsXRTFlux_11hrs}
\end{figure}

The comparison of the counting and likelihood techniques presented in Figure \ref{Fig:LikelihoodVsCounting_v2} reveals that the two methods produce consistent results when applied over short time intervals, but diverge when considering longer integration periods.  This is likely due to the increased sensitivity of the likelihood method, which takes into consideration the energy of the arriving photons, their position within the ROI, and the instrument PSF as function of energy.  High-energy photons that are consistent with the position of the candidate point source may only constitute a small percentage of the observed source flux over the expected background, but their arrival may be highly improbable (and hence significant) given the spectral shape of the assumed background model.  The likelihood analysis inherently accounts for this when calculating the significance of the source detection and may explain the diverging results returned by the two methods for long integration intervals.  As a result, we conclude that the joint likelihood method is more sensitive for source detection on intermediate and long timescales, but emphasize that the model assumptions and computational overhead inherent to the likelihood method still leave room for a counting analysis when searching for uncharacterized emission over a wide parameter space. 

The cumulative likelihood analysis presented in Figures \ref{Fig:TSVsPeakFlux} and \ref{Fig:TSVsXRTFlux_11hrs} reveals that both a burst's prompt gamma-ray flux and afterglow X-ray flux strongly correlate with the strength of the subthreshold emission.  These results are consistent with the interpretation that an extension of both the prompt and afterglow spectra contribute to the emission observed in the LAT energy range.  An inspection of Figure \ref{Fig:TSVsPeakFlux} reveals, though, a number of bright BAT-detected GRBs that nonetheless do not contribute significantly to the stacked signal significance.  These bursts are consistent with the population of bright GBM-detected bursts first reported in \citet{Ackermann2012} that are in the LAT FOV at the time of trigger but which produce no significant emission at MeV and GeV energies.  A detailed broadband spectral analysis of these LAT dark bursts by \citet{Ackermann2012} attributes their non-detection to spectral curvature of their high-energy spectrum, possibly due to pair attenuation.  

In contrast, the X-ray afterglow flux measured by XRT at 11 hrs correlates particularly strongly with the signal significance, with less than half the sample resulting in almost all of the observed excess.  This suggests that the prompt and afterglow phases may not both contribute to the LAT-detected emission for individual bursts, with the prompt phase contribution being suppressed in some cases.  The differences between Figures \ref{Fig:TSVsPeakFlux} and \ref{Fig:TSVsXRTFlux_11hrs} could imply that the LAT detection of the afterglow contribution could be due to simple threshold effects, whereas the detection of the prompt phase may be due to both threshold effects and an intrinsic suppression of the high-energy emission in some cases.  Detailed broadband spectral fits of XRT derived spectra and LAT upper limits will be required to decipher whether the LAT nondetections of the high-energy component of bright X-ray afterglows are simply due to instrumental sensitivity, or whether a break between the XRT and LAT energy ranges is required to explain the LAT nondetections.  Nonetheless, Figure \ref{Fig:TSVsXRTFlux_11hrs} shows that the X-ray afterglow flux is a strong predictor of the strength of the subthreshold LAT emission. 

Finally, we note that the TS value for the integration period covering the 50 s prior to the co-aligned trigger presented in Figure \ref{Fig:MaxTSVsTime}, while statistically not significant, remains elevated in comparison to the composite likelihood analysis performed over the same interval during the background orbits.  This may just reflect the varying trigger times with respect to the true start of the prompt emission, or it may suggest the presence of high-energy precursor emission prior to the prompt emission at keV energies.  Such activity has not yet been observed at MeV or GeV energies in any of the LAT-detected GRBs \citep{Ackermann2013}.  Ultimately, an analysis of a larger sample of bursts and/or the use of the upcoming Pass 8 event reconstruction \citep{Atwood2013}, which significantly improves the LAT sensitivity at low energies, will be needed to investigate whether such emission exists and how it compares temporally to activity at keV energies.

\section{Conclusions} \label{sec:Conclusions}

We perform a comprehensive stacking analysis of LAT data of \Swift~localized GRBs that were not detected by the LAT, but which fell within the instrument's FOV at the time of trigger.  We examine a total of 79 GRBs by comparing the observed counts over a range of time intervals to that expected from designated background orbits, as well as by using a joint likelihood technique to model the expected distribution of stacked counts, and find strong evidence for subthreshold emission at MeV to GeV energies using both techniques.  This observed excess is detected during intervals that include and exceed the durations typically characterizing the prompt emission observed at keV energies and lasts at least 2700 s after the co-aligned burst trigger.  

By utilizing a novel cumulative likelihood analysis, we are also able to identify which bursts contribute most significantly to the stacked signal.  We find that although a burst's prompt gamma-ray flux and afterglow X-ray flux both correlate with the strength of the subthreshold emission, the X-ray afterglow flux measured by XRT at 11 hrs post trigger correlates far more significantly. This suggests that although the prompt and afterglow phases may both contribute to the LAT-detected emission for individual bursts, the prompt phase contribution may be suppressed in some cases.  This is consistent with the population of bright GBM-detected bursts that are in the LAT FOV at the time of trigger, but which produce no significant emission at MeV energies and above.  

Overall, the extended nature of the subthreshold emission and its connection to the burst's afterglow brightness lend further support to the external forward shock origin of the late-time emission detected by the LAT.  These results suggest that the extended high-energy emission observed by the LAT may be a relatively common feature but remains undetected in a majority of bursts owing to instrumental threshold effects.

\acknowledgements
The \Fermi\ LAT Collaboration acknowledges generous ongoing support
from a number of agencies and institutes that have supported both the
development and the operation of the LAT as well as scientific data analysis.
These include the National Aeronautics and Space Administration and the
Department of Energy in the United States, the Commissariat \`a l'Energie Atomique
and the Centre National de la Recherche Scientifique / Institut National de Physique
Nucl\'eaire et de Physique des Particules in France, the Agenzia Spaziale Italiana
and the Istituto Nazionale di Fisica Nucleare in Italy, the Ministry of Education,
Culture, Sports, Science and Technology (MEXT), high-energy Accelerator Research
Organization (KEK) and Japan Aerospace Exploration Agency (JAXA) in Japan, and
the K.~A.~Wallenberg Foundation, the Swedish Research Council and the
Swedish National Space Board in Sweden.

Additional support for science analysis during the operations phase is gratefully
acknowledged from the Istituto Nazionale di Astrofisica in Italy and the Centre National d'\'Etudes Spatiales in France.

\begin{deluxetable}{lrrrrrrrrrrrrrr}
\tablecolumns{6} \tablewidth{0pc}
\tablecaption{Burst Sample}
\tabletypesize{\small}
\tablehead{ \colhead{GRB} & \colhead{Swift Trigger} &  \colhead{UTC} & \colhead{MET\tablenotemark{\dag}} & \colhead{RA} & \colhead{Dec}  & \colhead{Error} & \colhead{Source\tablenotemark{\ddag}}
\\
\colhead{} & \colhead{} &  \colhead{} &  \colhead{~s~} & \colhead{hr}  &\colhead{~$^{\circ}$~} & \colhead{$''$} & \colhead{}

}

\startdata

080905A &  323870 & 11:58:54 & 242308735 & 19:10:41.73 & --18:52:47.3 & 0.6 & G \\
080906A &  323984 & 13:33:16 & 242400797 & 15:12:10.65 & --80:31:03.2 & 0.6 & U \\
080916B &  324907 & 14:44:47 & 243269088 & 10:54:39.78 & +69:03:57.9 & 0.8 & U \\
080928A &  326115 & 15:01:32 & 244306893 & 06:20:16.84 & --55:11:58.9 & 0.5 & U \\
081003A &  20085 & 13:46:12 & 244734373 & 17:29:33.63 & +16:34:13.6 & 2.3 & X \\
081008A &  331093 & 19:58:09 & 245188690 & 18:39:50.00 & --57:25:52.0 & 0.6 & U \\
081012A &  331475 & 13:10:23 & 245509824 & 02:00:48.17 & --17:38:17.2 & 1.8 & X \\
081016B &  331856 & 19:47:14 & 245879235 & 00:58:15.44 & --43:31:48.5 & 1.8 & X \\
081029A &  332931 & 01:43:56 & 246937437 & 23:07:05.35 & --68:09:19.8 & 0.5 & U \\
081101A &  333320 & 11:46:31 & 247232792 & 06:23:20.40 & --00:06:18.0 & 102.0 & B \\
081102A$^{\beta}$ &  333427 & 17:44:39 & 247340680 & 22:04:41.58 & +52:59:39.9 & 1.5 & X \\
081104A &  333666 & 09:34:42 & 247484083 & 06:41:57.13 & --54:43:11.6 & 1.6 & X \\
081109B &  334129 & 13:47:16 & 247931237 & 23:20:31.68 & --55:54:43.2 & 204.0 & B \\
081118A &  334877 & 14:56:36 & 248712997 & 05:30:22.18 & --43:18:05.3 & 0.5 & G \\
081126A &  335647 & 21:34:10 & 249428051 & 21:34:03.59 & +48:42:38.3 & 0.5 & U \\
081127A &  335715 & 07:05:08 & 249462309 & 22:08:15.45 & +06:51:02.3 & 1.6 & X \\
081203B$^{\beta}$ &  20091 & 13:51:58 & 250005432 & 15:15:11.67 & +44:25:42.9 & 0.5 & U \\
081204A &  20092 & 16:44:55 & 250101896 & 23:19:09.45 & --60:13:29.6 & 4.3 & X \\
081222A &  337914 & 04:53:59 & 251614440 & 01:30:57.59 & --34:05:41.5 & 0.5 & U \\
081226B &  20095 & 12:13:11 & 251986392 & 01:41:58.80 & --47:26:20.4 & 150.0 & O \\
090113A &  339852 & 18:40:39 & 253564841 & 02:08:13.80 & +33:25:42.3 & 1.5 & X \\
090117A &  20097 & 15:21:54 & 253898516 & 10:56:10.48 & --58:14:00.6 & 2.2 & X \\
090129A &  341504 & 21:07:15 & 254956037 & 17:56:25.20 & --32:47:34.8 & 42.0 & B \\
090407A &  348650 & 10:28:25 & 260792907 & 04:35:55.07 & --12:40:45.1 & 1.4 & X \\
090422A &  349931 & 03:35:16 & 262064118 & 19:38:59.90 & +40:23:03.2 & 1.4 & X \\
090516A &  352190 & 08:27:50 & 264155272 & 09:13:02.62 & --11:51:15.4 & 0 & G \\
090518A &  352420 & 01:54:44 & 264304486 & 07:59:49.10 & +00:45:33.6 & 1.4 & X \\
090519A &  352648 & 21:08:56 & 264460138 & 09:29:07.00 & +00:10:49.1 & 0.5 & G \\
090529A$^{\beta}$ &  353540 & 14:12:35 & 265299157 & 14:09:52.54 & +24:27:32.2 & 0.7 & U \\
090621A &  355303 & 04:22:43 & 267250965 & 00:44:05.12 & +61:56:27.9 & 1.4 & X \\
090702A &  20106 & 10:40:37 & 268224039 & 11:43:35.39 & +11:30:06.5 & 5.5 & X \\
090708A &  356776 & 03:38:15 & 268717097 & 10:18:32.40 & +26:36:43.2 & 102.0 & B \\
090709B &  356912 & 15:07:42 & 268844864 & 06:14:01.44 & +64:04:26.4 & 96.0 & B \\
090712A &  357072 & 03:51:05 & 269063467 & 04:40:22.56 & +22:31:30.0 & 96.0 & B \\
090728A &  358574 & 14:45:45 & 270485147 & 01:58:36.60 & +41:37:59.6 & 0 & G \\
090813A$^{\alpha}$ &  359884 & 04:10:43 & 271829445 & 15:03:08.48 & +88:34:05.5 & 0.3 & G \\
090831C &  361489 & 21:30:25 & 273447027 & 07:13:10.63 & --25:07:07.2 & 1.5 & X \\
091127A$^{\alpha}$ &  377179 & 23:25:45 & 281057147 & 02:26:19.89 & --18:57:08.6 & 0.5 & G \\
091202A &  20123 & 23:10:12 & 281488214 & 09:15:19.75 & +62:32:59.0 & 0.5 & G \\
091208B$^{\alpha}$ &  378559 & 09:49:57 & 281958599 & 01:57:34.09 & +16:53:22.8 & 0.6 & U \\
091221A &  380311 & 20:52:52 & 283121574 & 03:43:11.40 & +23:14:28.3 & 0.6 & U \\
100111A &  382399 & 04:12:49 & 284875971 & 16:28:11.60 & +15:33:02.3 & 0.5 & U \\
100203A &  411011 & 18:31:07 & 286914669 & 06:24:54.00 & +04:47:34.8 & 120.0 & B \\
100206A$^{\alpha}$ &  411412 & 13:30:05 & 287155807 & 03:08:39.03 & +13:09:25.3 & 3.3 & X \\
100212A &  412081 & 14:07:22 & 287676444 & 23:45:40.23 & +49:29:40.7 & 1.0 & G \\
100316D &  416135 & 12:44:50 & 290436292 & 07:10:30.54 & --56:15:20.0 & 0.5 & G \\
100322B &  20129 & 07:06:18 & 290934380 & 05:05:57.36 & +42:41:06.0 & 240.0 & B \\
100401A &  20133 & 07:07:32 & 291798454 & 19:23:15.12 & --08:15:25.2 & 120.0 & B \\
100418A &  419797 & 21:10:08 & 293317810 & 17:05:27.09 & +11:27:42.3 & 0.5 & G \\
100427A &  20137 & 08:31:55 & 294049917 & 05:56:41.04 & --03:27:39.6 & 150.0 & B \\
100514A &  421962 & 18:53:58 & 295556040 & 21:55:17.46 & +29:09:36.1 & 0.3 & G \\
100528A$^{\alpha}$ &  20139 & 01:48:05 & 296704087 & 20:44:33.91 & +27:48:23.8 & 0.6 & U \\
100614A &  424716 & 21:38:26 & 298244308 & 17:33:59.85 & +49:14:02.9 & 1.7 & X \\
100704A &  426722 & 03:35:08 & 299907310 & 08:54:33.96 & --24:12:10.1 & 1.4 & X \\
100719A &  429357 & 03:30:57 & 301203059 & 07:29:16.56 & --05:51:25.2 & 144.0 & B \\
100725B &  429969 & 11:24:34 & 301749876 & 19:20:08.01 & +76:57:23.2 & 1.4 & X \\
100728B &  430172 & 10:31:55 & 302005917 & 02:56:13.47 & +00:16:52.2 & 0.5 & U \\
100802A$^{\beta}$ &  430603 & 05:45:36 & 302420738 & 00:09:52.38 & +47:45:18.8 & 0.4 & G \\
100902A &  433160 & 19:31:53 & 305148715 & 03:14:30.96 & +30:58:44.8 & 0.5 & G \\
101129A &  20151 & 15:39:31 & 312737973 & 10:23:41.04 & --17:38:42.0 & 180.0 & B \\
101219B &  440635 & 16:27:53 & 314468875 & 00:48:55.35 & --34:33:59.5 & 0.6 & U \\
110102A$^{\alpha}$ &  441454 & 18:52:25 & 315687147 & 16:23:31.41 & +07:36:49.8 & 0.5 & U \\
110107A &  20154 & 21:15:51 & 316127753 & 19:59:38.23 & +41:54:51.4 & 3.0 & X \\
110112B &  20155 & 22:24:55 & 316563897 & 00:42:23.76 & +64:24:21.6 & 156.0 & O \\
110128A &  443861 & 01:44:33 & 317871875 & 12:55:35.10 & +28:03:54.1 & 0 & G \\
110223A &  446674 & 20:56:59 & 320187421 & 23:03:24.52 & +87:33:28.3 & 2.0 & X \\
110411A &  451165 & 19:34:11 & 324243253 & 19:25:45.65 & +67:42:39.1 & 1.0 & G \\
110412A &  451191 & 07:33:21 & 324286403 & 08:53:57.84 & +13:29:16.8 & 114.0 & B \\
110414A &  451343 & 07:42:14 & 324459736 & 06:31:29.50 & +24:21:44.7 & 1.4 & X \\
110801A &  458521 & 19:49:42 & 333920984 & 05:57:44.87 & +80:57:21.3 & 0.5 & U \\
110808A$^{\beta}$ &  458918 & 06:18:54 & 334477136 & 03:49:04.27 & --44:11:40.1 & 0.6 & U \\
110825A$^{\alpha}$  &  20183 & 02:26:50 & 335932012 & 02:59:35.04 & +15:24:25.2 & 7115.0 & O \\
110903A$^{\beta}$ &  20184 & 02:39:55 & 336710397 & 13:08:15.82 & +58:58:53.8 & 1.6 & X \\
111029A &  506519 & 09:44:40 & 341574282 & 02:59:08.07 & +57:06:39.5 & 1.9 & X \\
111117A$^{\alpha}$ &  507901 & 12:13:41 & 343224823 & 00:50:46.26 & +23:00:40.0 & 0.1 & O \\
111204A &  509018 & 13:37:28 & 344698650 & 22:26:30.81 & --31:22:29.3 & 1.9 & X \\
111208A &  20190 & 08:28:11 & 345025693 & 19:20:53.84 & +40:40:34.2 & 1.6 & X \\
111212A &  509543 & 09:23:07 & 345374589 & 20:41:43.57 & --68:36:44.7 & 1.4 & X \\
120118B &  512003 & 17:00:21 & 348598823 & 08:19:29.04 & --07:11:05.1 & 1.4 & X \\

\enddata
\tablenotetext{\dag}{Mission elapsed time relative to January 1, 2001, 0h:0m:0s UTC}
\tablenotetext{\ddag}{B = \emph{Swift} BAT, U = \emph{Swift} UVOT, X = \emph{Swift} XRT, G = Ground, O = Other Missions}
\tablenotetext{\alpha}{GRBs that triggered an autonomous repoint request (ARR) of the \Fermi~spacecraft}
\tablenotetext{\beta}{GRBs that occurred while the \Fermi~spacecraft was performing a target of opportunity (ToO) observation}

\label{Table:SampleDefinition} 
\end{deluxetable}

\bibliography{ms}

\end{document}